\documentclass[prx,twocolumn,groupedaddress]{revtex4-2}
\usepackage{xcolor}
\usepackage{amsfonts,amsmath,amssymb}
\usepackage{commath}
\usepackage{textgreek}
\usepackage{todonotes}
\usepackage{lmodern}
\usepackage[T1]{fontenc}
\usepackage[utf8x]{inputenc}
\usepackage[UKenglish]{babel}
\usepackage{times,inconsolata}
\usepackage{microtype,bbm}

\usepackage[colorlinks = true,
            linkcolor = blue,
            urlcolor  = blue,
            citecolor = blue,
            anchorcolor = blue]{hyperref}

\usepackage[capitalize]{cleveref}
\usepackage{graphicx}
\usepackage{xcolor}
\graphicspath{{ms_pub_figs/}}

\begin{document}

\title{Generation of Photonic Matrix Product States with Rydberg Atomic Arrays}

\author{Zhi-Yuan Wei$^{1,2}$, Daniel Malz$^{1,2}$, Alejandro Gonz\'alez-Tudela$^{3}$, J. Ignacio Cirac$^{1,2}$}
\affiliation{$^1$ Max-Planck-Institut f{\"u}r Quantenoptik, Hans-Kopfermann-Stra{\ss}e 1, D-85748 Garching, Germany}

\affiliation{$^2$Munich Center for Quantum Science and Technology (MCQST), Schellingstr. 4, D-80799 M{\"u}nchen, Germany}

\affiliation{\textit{$^3$ Instituto de F\'isica Fundamental IFF-CSIC, Calle Serrano 113b, Madrid 28006, Spain}}

\date{\today}

\begin{abstract}
We show how one can deterministically generate photonic matrix product states with high bond and physical dimensions with an atomic array if one has access to a Rydberg-blockade mechanism. We develop both a quantum gate and an optimal control approach to universally control the system and analyze the photon retrieval efficiency of atomic arrays. Comprehensive modeling of the system shows that our scheme is capable of generating a large number of entangled photons. We further develop a multi-port photon emission approach that can efficiently distribute entangled photons into free space in several directions, which can become a useful tool in future quantum networks.
\end{abstract}

\maketitle

\section{Introduction}

The generation of multiphoton quantum states is at the core of many quantum technologies, including computing~\cite{OBrien2009}, cryptography~\cite{gisin2007quantum}, networks~\cite{kimble2008quantum}, or sensing~\cite{Degen2017}. The standard method utilizes parametric down-conversion (PDC)~\cite{burnham1970observation}, where the desired state is generated by heralding its presence by post-selection. While very successful in several scenarios, this method possesses certain limitations, notably the exponential dependence of the success probability on the number of photons. These limitations can be overcome, for instance, by using emitters coupled to photonic waveguides, where one first prepares an entangled state of the emitters, which is then mapped into the multiphoton state using collective decay~\cite{Gonzalez-Tudela2015,PhysRevLett.118.213601}. This requires an effective non-linear interaction (coherent or dissipative) among the emitters, and the number of photons is limited by the number of emitters. An alternative way of generating multiphoton states at a higher rate is to do it sequentially using a photon source with long coherence time~\cite{gheri1998entanglement,saavedra2000controlled,schon2007,Lindner2009,Nielsen2010,Tiurev2020a}. The class of states that can be generated in this way coincides with the set of matrix product states (MPS)~\cite{schon2005}, a class of tensor network states that have been extensively studied in condensed matter physics~\cite{Perez-Garcia2006,schollwock2011density,orus2014practical,Orus2019}. Some of the sequential generation protocols have recently been experimentally realized in quantum dots~\cite{Lindner2016} and circuit QED~\cite{Eichler2015,Besse2020}.
Going beyond one-dimensional MPS, it has been proposed to produce a subclass of projected entangled pair states (PEPS)~\cite{verstraete2004renormalization,verstraete2008matrix,Orus2019}
by allowing the emitted photons to travel back and interact with the photon source again~\cite{Pichler2017, Dhand2018, Xu2018}, coupling multiple MPS sources~\cite{Economou2010,Gimeno-Segovia2019,Russo_2019,Bekenstein}, or applying linear optical operations on certain resource states \cite{Lubasch2018}.

Most of the above-mentioned schemes produce tensor network states of photonic qubits with a limited bond dimension (typically, $D = 2$), which contains important states like the GHZ state~\cite{greenberger1989going} and the cluster state~\cite{Briegel2001}. A notable exception is the scheme studied in Ref.~\cite{Dhand2018}, which is capable of probabilistically producing tensor network states with small $D>2$ utilizing the PDC process in an optical loop. Nevertheless, It is very desirable to deterministically produce tensor network states with moderately higher bond dimensions, as the ground states of a large variety of many-body systems are encoded there~\cite{orus2014practical, Huang2015a,Dalzell2019,Huang2019a,Schuch2017}. Tensor network states of high bond dimensions can also be instrumental for metrological purposes \cite{Jarzyna2013, Chabuda2020}. Moreover, tensor network states with higher physical dimensions (photonic qudit) further allows one to explore high-dimensional entanglement~\cite{Erhard2020} with many applications, for instance, to produce bosonic quantum error-correcting codes~\cite{Michael2016}, or act as high-dimensional resource states for state preparation~\cite{ContrerasTejada2019a} and computation~\cite{Zhou2003, Verstraete2004c,clark2006valence, Wang2017,Wei2018}.

A very attractive way of generating photonic states is by using atomic ensembles and arrays in free space~\cite{Saffman2002,Porras2008,Pedersen2009,Dudin2012,Li2013a,Miroshnychenko2013,Feng2014,Petrosyan2018}. The most significant advantage is that that system does not require the presence of a cavity or a waveguide to collect the photons; interference effects are harnessed instead to emit the photons in a determined direction with a high probability, a phenomenon that is enhanced by collective effects (superradiance) in three spatial dimensions. In Ref.~\cite{Porras2008} a scheme was proposed to efficiently generate photonic states in atomic arrays. The basic idea is to combine those interference/collective phenomena with a Rydberg blockade mechanism~\cite{jaksch00a,Lukin2001a}, so that in a first step an entangled atomic state is prepared, which is mapped in a second step into photonic degrees of freedom. This results in one or two outgoing photonic wavepackets in a state that is predetermined by the procedure.

In this paper we show how, in a similar setup, it is possible to generate quantum photonic states in a sequential way, where the output states consist of many wavepackets outgoing in predetermined directions and whose quantum state is an arbitrary MPS. This generalizes the class of states that can be produced on a similar setup~\cite{Nielsen2010} (a subclass of MPS with $D=2$) with a simpler atomic level scheme. We illustrate the method with simple examples, like the cluster state~\cite{Briegel2001} (an MPS with $D=2$) and the generalized GHZ states~\cite{ContrerasTejada2019a} (MPS with high bond dimension and physical dimension). We analyze the performance of the source by careful simulation of the impact of imperfections in this scheme and characterizing the photon retrieval efficiencies of atomic arrays with various geometries. We predict that with this scheme and realistic conditions it should be possible to generate several tens of entangled photons, and extend the setup to send the photons to different directions by controlling the phase-matching direction of the collective excitations. Thus, this setup provides us with an excellent platform for the sequential generation of photonic MPS, which can efficiently distribute the entanglement among multiple ports.

The rest of the manuscript is structured as follows. In \cref{seq_gen}, we present our scheme to generate arbitrary MPS using a Rydberg-blockaded atomic array. In \cref{unit_imple}, we develop both a quantum gate approach and a quantum optimal control approach~\cite{DAlessandro2007,Werschnik2007} to implement arbitrary unitaries in our system, illustrate their usage by constructing unitaries required for generating the cluster state and the generalized GHZ state, and analyze the impact of imperfections during the unitary evolution process. In \cref{pho_ret_main}, we characterize the photon retrieval efficiencies of atomic arrays with various geometries and the impact of imperfections during the photon emission process. Then we compute the number of entangled photons that can be created with our scheme in \cref{tot_pho_opt} and extend it to operate as a multi-port device in \cref{fse}. We propose a possible experimental realization in \cref{rb_imple} and summarize our work in \cref{sum_olk}.

\section{MPS generation with a Rydberg array}
\label{seq_gen}
The general protocol for the sequential generation of MPS is presented in Ref.~\cite{schon2005}. In this section, we first briefly review this protocol in \cref{mps_prot_gen}, and then introduce our Rydberg-blockaded atomic array setup in \cref{sys_mb}.

\subsection{MPS generation protocol}
\label{mps_prot_gen}
Let us consider a photon source with Hilbert space ${\cal H}_{{\rm{src}}}^{\left( {D,d} \right)}$ consisting of a $D$-dimensional ancilla with Hilbert space ${\cal H}_D$ coupled to a $d$-dimensional emitter with Hilbert space ${\cal H}_d$, that ${{\cal H}^{\left( {D,d} \right)}_{{\rm{src}}}} = {{\cal H}_D} \otimes {{\cal H}_d}$. We assume that we can implement any arbitrary unitary operation ${U_{\rm src}} \in \rm{SU}({D \cdot d})$ in ${{\cal H}_{{\rm{src}}}^{\left( {D,d} \right)}}$, and we can trigger a photon emission process ${M_P}:{{\cal H}_d} \to {{\cal H}_d} \otimes {{\cal H}_{{\rm{\rm ph}}}}$ that generates a $d$-dimensional photonic qudit with Hilbert space ${{\cal H}_{\mathrm{\rm ph}}}$:
\begin{equation} \label{ap_map}
{M_P}:{\left| i \right\rangle _d} \to {\left| 0 \right\rangle _d}{\left| i \right\rangle _{{\rm{\rm ph}}}},\qquad i \in \left( {0,...,d - 1} \right).
\end{equation}

The sequential photon generation protocol starts with an initial state ${\left| {{\varphi _I}} \right\rangle _D}{\left| 0 \right\rangle _d}$ without excitations on the emitter. In each round, we first apply a unitary on the photon source ${U_{\left[ i \right]}} \in {{\cal H}_{{\rm{src}}}^{\left( {D,d} \right)}}$ and then trigger a photon emission $M_P$, which produces a photonic qudit. Thus, in this protocol, the unitaries always act on states of form ${\left| \varphi  \right\rangle _D}{\left| 0 \right\rangle _d}$ as
\begin{equation} \label{u_act}
{U_{\left[ i \right]}}{\left| \varphi  \right\rangle _D}{\left| 0 \right\rangle _d} = \sum\limits_{{j} = 0}^{d - 1} {V_{\left[ i \right]}^{{j}}} {\left| \varphi  \right\rangle _D}{\left| j \right\rangle _d}.
\end{equation}
After operating the protocol for $n$ rounds (shown in \cref{fig1}(b)), we get the final state
\begin{equation} \label{sysph_state}
\left| \Psi  \right\rangle  = {M_P}{U_{\left[ n \right]}}...{M_P}{U_{\left[ 1 \right]}}\left| {{\varphi _I}} \right\rangle .
\end{equation}
Assuming that after the $n$-th photon emission, the photon source disentangles from the photonic state, such that $\left| \Psi  \right\rangle  = {\left| {{\varphi _F}} \right\rangle _D}{\left| 0 \right\rangle _d} \otimes \left| {{\psi _{{\rm{MPS}}}}} \right\rangle $, we get a $n$-qudit photonic MPS of the form~\cite{schon2005} (see \cref{apd_sup} for details)
\begin{equation} \label{mps}
\left| {{\psi _{{\rm{MPS}}}}} \right\rangle _{\rm ph}  = \sum\limits_{{i_1},...,{i_n} = 0}^{d - 1} {_D\left\langle {{\varphi _F}} \right|V_{\left[ n \right]}^{{i_n}}...V_{\left[ 1 \right]}^{{i_1}}{{\left| {{\varphi _I}} \right\rangle }_D}{{\left| {{i_n}...{i_1}} \right\rangle }_{{\rm{\rm ph}}}}} .
\end{equation}
If one is able to generate arbitrary unitaries in ${{\cal H}^{\left( {D,d} \right)}_{{\rm{src}}}}$, then the $D$-dimensional matrices $\{ {V_{\left[ k \right]}^{{i_k}}} \}$ can be arbitrary, as long as the isometry condition $\sum\nolimits_{{i_k} = 0}^1 {V_{\left[ k \right]}^{{i_k}\dag }V_{\left[ k \right]}^{{i_k}}}  = {I_D}$ is fulfilled~\cite{schon2005}. This shows that in principle, the whole family of MPS can be generated in this way. Furthermore, by including another $m$-level ancilla, it is possible to create superposition of the $m$-level ancilla and photonic MPSs of the form (see \cref{apd_sup} for details)
\begin{equation} \label{mps_sup}
\left| {{\psi _{{m,\rm{ph}}}}} \right\rangle  = \sum\limits_{i = 0}^{m - 1} {{\alpha _i}{{\left| i \right\rangle }_{{m}}}{{\left| {\psi _{{\rm{MPS}}}^i} \right\rangle }_{{\rm{\rm ph}}}}}.
\end{equation}
This additional ancilla can be either a physical system or a part of the Hilbert space of a high-dimensional ancilla. Such superposed states find many applications in quantum networks \cite{Miguel-Ramiro2020a}, and realize the functionality of quantum random access memory \cite{Giovannetti2008}.

A good physical platform to generate MPS of high bond and physical dimensions with the above protocol requires one to
\begin{enumerate}
	\item have a Hilbert space or its subspace of structure ${\cal H}_{{\rm{src}}}^{\left( {D,d} \right)}$ with high dimensions,
	\item be able to efficiently implement unitary operations [cf.~\cref{u_act}],
	\item be able to realize the photon emission $M_P$ [cf.~\cref{ap_map}].
\end{enumerate}
In the following subsections we show that a Rydberg blockaded~\cite{jaksch00a,Lukin2001a} atomic array is well-suited for this task.

\subsection{System setup}
\label{sys_mb}
\begin{figure*}
    \centering
    \includegraphics[width=\textwidth]{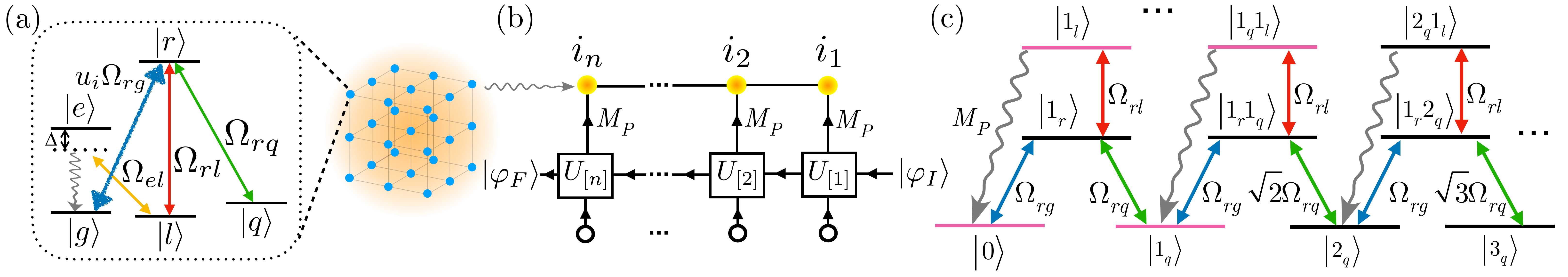}
        \caption{Generation of photonic MPS with a Rydberg-blockaded array. (a) Schematic plot and single-atom level scheme of our protocol. We consider an atomic array within the Rydberg blockade radius. For each atom, there are three hyperfine ground states $\left| g \right\rangle ,\left| l \right\rangle ,\left| q \right\rangle $ coupled to an intermediate Rydberg state $\left| {{r}} \right\rangle $ via lasers denoted by arrows of different colours. A fast decaying excited state $\left| e \right\rangle $ is used to convert the collective excitation(s) on $\left| l \right\rangle $ to photon(s). (b) The sequential photon generation protocol. Starting from the atomic initial state $\left| {{\varphi _I}} \right\rangle $, the photonic state is generated by successively applying the unitary operation ${U_{\left[ i \right]}}$ followed by the photon emission process $M_P$ for $n$ times, which yields an $n$-photon matrix product state. (c) The structure of collective Hilbert space ${\mathcal{H}_{\rm col}}$ in the low-excitation regime (without $\left| e \right\rangle$), with the basis defined in \cref{hcol_basis}. The arrows indicate the same laser transitions with the same coloring as in (a). The subspace used in our scheme for generating the cluster state is shown in pink (see \cref{unit_imple}).}
        
        \label{fig1}
\end{figure*}

We consider an array of $N$ atoms (sketched in \cref{fig1}(a)), with three hyperfine ground states $\left| {g} \right\rangle $, $\left| {l} \right\rangle $ and $\left| {q} \right\rangle $, a Rydberg state $\left| {r} \right\rangle $, and an auxiliary excited state $\left| e \right\rangle $ used for the photon emission. The transitions between all these states are controlled with lasers (denoted with coloured arrows in \cref{fig1}(a)). We assume our system is inside the Rydberg blockade radius, such that at most one Rydberg excitation can be present in the system. In addition, the system is operated in the \textit{low-excitation} regime, which means that most atoms are in the ground state $\left| g \right\rangle $. We denote the many-body ground state of the system by $\left| 0 \right\rangle  = {\left| g \right\rangle ^{ \otimes N}} $.

Let us first consider the unitary evolution process in our protocol under ideal conditions, which means (i) the Rydberg blockade condition is perfect, and (ii) there are no decoherence processes. Since the state $\left| e \right\rangle $ is not involved in the unitary operations, we define $\sigma _{\alpha \beta }^i = {\left| \alpha  \right\rangle _i}\left\langle \beta  \right|$ and write the Hamiltonian for our atomic array as follows
\begin{equation} \label{h_tot}
    H\left( t \right) = {H_0} + {H_{rg}}\left( t \right) + {H_{rl}}\left( t \right) + {H_{rq}}\left( t \right),
\end{equation}
where ${H_0}$ contains the energy of the atomic levels (with $\hbar  = 1$)
\begin{equation}
{H_0} = {\omega _r}\sum\limits_{i = 1}^N {\sigma _{rr}^i}  + {\omega _q}\sum\limits_{i = 1}^N {\sigma _{qq}^i}  + {\omega _l}\sum\limits_{i = 1}^N {\sigma _{ll}^i},
\end{equation}
and we have three driving lasers, denoted by $\mathrm{L_{g}}$, $\rm{L_{q}}$ and $\mathrm{L_{l}}$, which are used to connect the levels $\left| g \right\rangle $, $\left| q \right\rangle $ and $\left| l \right\rangle $ with the Rydberg level $\left| r \right\rangle $ during the unitary process. We define the collective excitation operator 
\begin{equation} \label{exci_op_g}
	S_{\alpha ,{\bf{k}}}^\dag  = \sum\limits_{i = 1}^N {{u_{i}}{e^{i{\bf{k}} \cdot {{\bf{r}}_i}}}\sigma _{\alpha g}^i}, \qquad \alpha  = r,q,l,
\end{equation}
where the ${u_i}\sim O( {1/\sqrt N } )$ is the normalized laser profile for $\mathrm{L_{g}}$ with $\sum\nolimits_i {{{\left| {{u_{i}}} \right|}^2}}  = 1$. The Hamiltonian ${H_{rg}}\left( t \right)$ corresponding to the laser $\mathrm{L_{g}}$ is
\begin{equation}\label{hrg_ext}
{H_{rg}}\left( t \right) = \frac{1}{2}\left( {{\Omega _{rg}}\left( t \right){e^{ - i{\omega _{rg}}t}}S_{r,{{\bf{k}}_{rg}}}^\dag  + {\rm{H}}.{\rm{c}}.} \right).
\end{equation}
Starting from $\left| 0 \right\rangle $, ${H_{rg}}(t)$ leads to a Rabi evolution between $\left| 0 \right\rangle $ and $S_{r,{{\bf{k}}_{rg}}}^\dag \left| 0 \right\rangle $ with collective Rabi frequency ${\Omega _{rg}}\left( t \right)$. The lasers $\mathrm{L_{l}}$ and $\mathrm{L_{q}}$ have plane-wave profiles, with Hamiltonians
\begin{equation} \label{hlq_ext}
\begin{array}{l}
{H_{r\alpha }}\left( t \right) = \frac{1}{2}\sum\limits_{i = 1}^N {\left( {{\Omega _{r\alpha }}\left( t \right)\sigma _{r\alpha }^i{e^{i\left( {{{\bf{k}}_{r\alpha }} \cdot {{\bf{r}}_i} - {\omega _{r\alpha }}t} \right)}} + {\rm{H}}.{\rm{c}}.} \right)} ,\\
\qquad \alpha  = l,q.
\end{array}
\end{equation}

In the low-excitation regime, we can apply the Holstein-Primakoff approximation~\cite{Holstein1940} to our system to treat the collective excitation operators $S_{l,{\bf{k}}}^\dag $ and $S_{q,{\bf{k}}}^\dag $ as bosonic operators, whereas the Rydberg collective excitation operator $S_{r,{\bf{k}}}^\dag $ can be treated as a spin-1/2 creation operator due to the Rydberg blockade. We have
\begin{equation} \label{ops_approxi}
	S_{r,{{\bf{k}}_{rg}}}^\dag  \approx {\sigma _r^\dag},\quad S_{q,{{\bf{k}}_{rg}} - {{\bf{k}}_{rq}}}^\dag  \approx a_q^\dag ,\quad S_{l,{{\bf{k}}_{rg}} - {{\bf{k}}_{rl}}}^\dag  \approx a_l^\dag.
\end{equation}
In our MPS generation protocol, we fix the laser momenta of $\mathrm{L_{g},L_{q}}$ and $\mathrm{L_{l}}$, and hence omit the momentum dependence of the approximated operators (right hand side of \cref{ops_approxi}) for notational simplicity. The lasers $\mathrm{L_{g},L_{q}}$ and $\mathrm{L_{l}}$ couple states in a collective Hilbert space ${{\cal H}_{{\rm{col}}}}$ consisting of a qubit formed by presence or absence of a collective Rydberg excitation with Hilbert space ${\cal H}_r$, and two oscillators formed by collective excitations on $\left| q \right\rangle $ and $\left| l \right\rangle $ with Hilbert spaces ${\cal H}_q$ and ${\cal H}_l$, such that ${{\cal H}_{{\rm{col}}}} = {{\cal H}_r} \otimes {{\cal H}_q} \otimes {{\cal H}_l}$. A basis of ${{\cal H}_{{\rm{col}}}}$ is spanned by
\begin{equation} \label{hcol_basis}
\left| {{s_r}{m_q}{n_l}} \right\rangle  = \frac{1}{{\sqrt {{m_q}!{n_l}!} }}{( {\sigma _r^\dag } )^{{s_r}}}{( {a_q^\dag } )^{{m_q}}}{( {a_l^\dag } )^{{n_l}}}\left| 0 \right\rangle.
\end{equation}
Here, $s_r = 0,1$ and $m_q,n_l$ are positive integers. For notational simplicity, we further omit the label(s) of the mode(s) with no excitations, for instance we define $\left| {{s_r}{m_q}} \right\rangle  \equiv \left| {{s_r}{m_q}{0_l}} \right\rangle$. The structure of ${{\cal H}_{\rm col}}$ and the laser-induced transitions is depicted in \cref{fig1}(c).

The collective Hilbert space ${{\cal H}_{{\rm{col}}}}$ contains the the structure ${{\cal H}^{\left( {D,d} \right)}_{{\rm{src}}}}$ needed for the MPS generation protocol in \cref{mps_prot_gen}. We identify the first $D$ Fock states in ${\mathcal H}_q$ as a $D$-level ancilla, and the first $d$ Fock states in ${\mathcal H}_l$ as a $d$-level quantum emitter. Thus this platform fulfills the requirement 1 for a good MPS source listed in \cref{mps_prot_gen}.

Using the definitions of the collective operators [\cref{ops_approxi}], we can rewrite the system Hamiltonian [\cref{h_tot}] as
\begin{equation} \label{hc_tot}
H_{\rm col}\left( t \right) = H_0^{\rm col} + H_{rg}^{\rm col}\left( t \right) + H_{rl}^{\rm col}\left( t \right) + H_{rq}^{\rm col}\left( t \right),
\end{equation}
where
\begin{equation}
{H_0^{\rm col}} = {\omega _r}\sigma _r^\dag {\sigma _r} + {\omega _l}a_l^\dag {a_l} + {\omega _q}a_q^\dag {a_q},
\end{equation}
\begin{equation}
{H_{rg}^{\rm col}}\left( t \right) = \frac{{{\Omega _{rg}}\left( t \right)}}{2}\left( {{e^{ - i{\omega _{rg}}t}}\sigma _r^\dag  + {\rm{H}}.{\rm{c}}.} \right),    
\end{equation}
\begin{equation} \label{hlq_col}
{H_{r\alpha}^{\rm col}}\left( t \right) = \frac{{{\Omega _{r\alpha }}\left( t \right)}}{2}\left( {{e^{ - i{\omega _{r\alpha }}t}}{a_\beta }\sigma _r^\dag  + {\rm{H}}.{\rm{c}}.} \right)\;\; \alpha  = l,q.
\end{equation}
By combining the universal controllability for the Jaynes-Cummings model \cite{Yuan2007,Mischuck2013} and the Lemma 5.5 of Ref.~\cite{Hofmann2017}, one can show that any arbitrary unitary in ${{\cal H}_{\rm col}}$ can be generated with time evolution under $H_{\rm col}\left( t \right)$, and the optimal time cost to implement a unitary in $H_{\rm src}^{\left( {D,d} \right)}$ scales as ${T_{D,d}}\sim{\left( {Dd} \right)^2}$ (see \cref{dim_scale}). Thus this platform fulfills the requirement 2 listed in \cref{mps_prot_gen}. We also estimate the scaling of implementing the unitary as a function of the dimension in \cref{dim_scale}.

One can also engineer a on-demand photon emission process ${M_P}:{{\cal H}_l} \to {{\cal H}_l} \otimes {{\cal H}_{\rm ph}}$ by coupling $\left| l \right\rangle $ to an excited state $\left| e \right\rangle $ with a plane-wave profile laser ${\rm L}_{\rm e}$, which triggers the emission of the photon(s) \cite{Porras2008}
\begin{equation} 
{M_P}:\left| {{s_r}{m_q}{n_l}} \right\rangle  \to \left| {{s_r}{m_q}} \right\rangle {\left| {{n_l}} \right\rangle _{{\text{\rm ph}}}}
\end{equation}
This fulfills the requirement 3 listed in \cref{mps_prot_gen}. Here, we assume that the atomic population only goes from $\left| e \right\rangle $ to $\left| g \right\rangle $, which for instance can be realized by choosing $\left| e \right\rangle  \leftrightarrow \left| g \right\rangle$ as a cycling transition~\cite{Porras2008}. Such high-efficiency photon retrieval of a spin-wave stored in an atomic array has been demonstrated experimentally~\cite{Dudin2013,Yang2016}. More details of the photon retrieval process are presented in \cref{pho_ret_main}.

With all three requirements for an efficient MPS source fulfilled, this setup can be used to generate photonic MPS with high bond dimension $D$ and physical dimension $d$ satisfying $D + d -2 \ll N$ (required by the low-excitation regime) using the generic protocol introduced in \cref{mps_prot_gen}.

\section{Implementation of unitary operations}
\label{unit_imple}
The Hamiltonian written in the collective basis [\cref{hc_tot}] is equivalent to a central qubit coupled to two oscillators. Universal control of qubit-oscillator systems has been studied in various setups, in particular the Jaynes-Cummings model of a single qubit coupled to an oscillator. While state preparation can be done using the Law-Eberly scheme~\cite{Law1996}, efficiently implementing arbitrary unitary operations is more challenging~\cite{Santos2005,Strauch2012a,Mischuck2013,Krastanov2015,Fosel2020,Fosel2020}.
In this section, we present efficient unitary control methods for this setup. We first provide a quantum gate approach that utilizes the AC Stark shift to implement a set of universal gates needed for generating photonic MPS of bond dimension $D=2$ and physical dimension $d=2$ in \cref{stark_circ}. Then, we present a quantum optimal control scheme that achieves universal control in ${{\cal H}_{\rm col}}$ in \cref{opt_main}. We demonstrate these two approaches by constructing unitaries required for generating the cluster state~\cite{Briegel2001} and the generalized GHZ state~\cite{ContrerasTejada2019a}. We further analyze the impact of imperfections during the photon generation protocol in \cref{real_model}.

\subsection{Quantum gate approach}
\label{stark_circ}
The class of MPS of $D = 2$ and $d = 2$ contains many interesting states like the GHZ state~\cite{greenberger1989going} and the cluster state. To generate this class of states with our scheme, it is sufficient if both ancilla and emitter are two-level qubits. Thus, ${\cal H}^{\left( {2,2} \right)}_{{\rm{src}}} = {\rm{span}}\{ {\left| 0 \right\rangle ,\left| {{1_l}} \right\rangle ,\left| {{1_q}} \right\rangle ,\left| {{1_q}{1_l}} \right\rangle } \}$ (the basis of ${\cal H}^{\left( {2,2} \right)}_{{\rm{src}}}$ is depicted in pink in \cref{fig1}(c), and cf. \cref{hcol_basis} for the definition of basis).

The procedure to implement a unitary in ${\cal H}^{\left( {2,2} \right)}_{{\rm{src}}}$ consists of three steps.
\begin{enumerate}
	\item Apply a $\pi$-pulse of the driving $\mathrm{L_{l}}$ to transfer the population from $\left| l \right\rangle $ to the Rydberg level $\left| r \right\rangle $ as $\left| {{1_l}} \right\rangle  \to \left| {{1_r}} \right\rangle $ and $\left| {{1_q}{1_l}} \right\rangle  \to \left| {{1_r}{1_q}} \right\rangle $.
	\item Implement the desired operation within a space \[{\cal H}_{{\rm{mid}}} = {\rm{span}}\left\{ {\left| 0 \right\rangle ,\left| {{1_r}} \right\rangle ,\left| {{1_q}} \right\rangle ,\left| {{1_r}{1_q}} \right\rangle } \right\}.\] Here ${\cal H}_{{\rm{mid}}}$ is similar to ${\cal H}^{\left( {2,2} \right)}_{{\rm{src}}}$, but with the collective excitation(s) on $\left| l \right\rangle $ transformed to the Rydberg state $\left| r \right\rangle $.
	\item Apply another $\pi$-pulse with the laser $\mathrm{L_{l}}$ to transfer the population from $\left| r \right\rangle $ back to $\left| l \right\rangle $ as $\left| {{1_r}} \right\rangle  \to \left| {{1_l}} \right\rangle $ and $\left| {{1_r}{1_q}} \right\rangle  \to \left| {{1_q}{1_l}} \right\rangle $.
\end{enumerate}
The key step is to implement arbitrary unitaries in ${{\cal H}_{\rm mid}}$, for which we need to construct a universal two-qubit gate set in ${{\cal H}_{\rm mid}}$ using the lasers ${\rm L}_g$ and ${\rm L}_q$. $\mathrm{L_{g}}$ can realize arbitrary rotations in ${\cal H}_r$ that couples $\left| 0 \right\rangle  \leftrightarrow \left| {{1_r}} \right\rangle $ and $\left| {{1_q}} \right\rangle  \leftrightarrow \left| {{1_r}{1_q}} \right\rangle $, whereas $\mathrm{L_{q}}$ couples $\left| {{1_r}} \right\rangle  \leftrightarrow \left| {{1_q}} \right\rangle $. However, ${\rm L}_q$ also leads to population leakage from $\left| {1_r1_q} \right\rangle $ to $\left| {{2_q}} \right\rangle $, which is outside of ${{\cal H}_{\rm mid}}$.

To avoid such leakage error, we propose to use the AC Stark shift~\cite{Schmidt-Kaler2004,Lee2019a} induced by applying ${\rm L}_q$ with a large detuning $\Delta_{\rm ST} \gg {\Omega _{rq}}$. This allows us to construct a phase gate of the form ${S_T}\left( \theta  \right) =\textrm{diag}\left( {1,{e^{ - i\theta }},{e^{i\theta }},{e^{ - 2i\theta }},{e^{2i\theta }}} \right)$ with the basis $\{ {\left| 0 \right\rangle ,\left| {{1_r}} \right\rangle ,\left| {{1_q}} \right\rangle ,\left| {{1_r}{1_q}} \right\rangle ,\left| {{2_q}} \right\rangle } \}$. Using ${S_T}\left( \theta  \right)$ together with single-qubit rotations, we can construct the SWAP gate and CNOT gate in ${{\cal H}_{\rm mid}}$ (see \cref{circ_unit} for details). The SWAP gate, CNOT gate and the single-qubit rotations on ${{\cal H}_r}$ realize a universal gate set in ${{\cal H}_{\rm mid}}$, and thus lead to the universal control of ${\cal H}^{\left( {2,2} \right)}_{{\rm{src}}}$ using the quantum gate approach.

\begin{figure}
    \centering
    \includegraphics[width=0.49\textwidth]{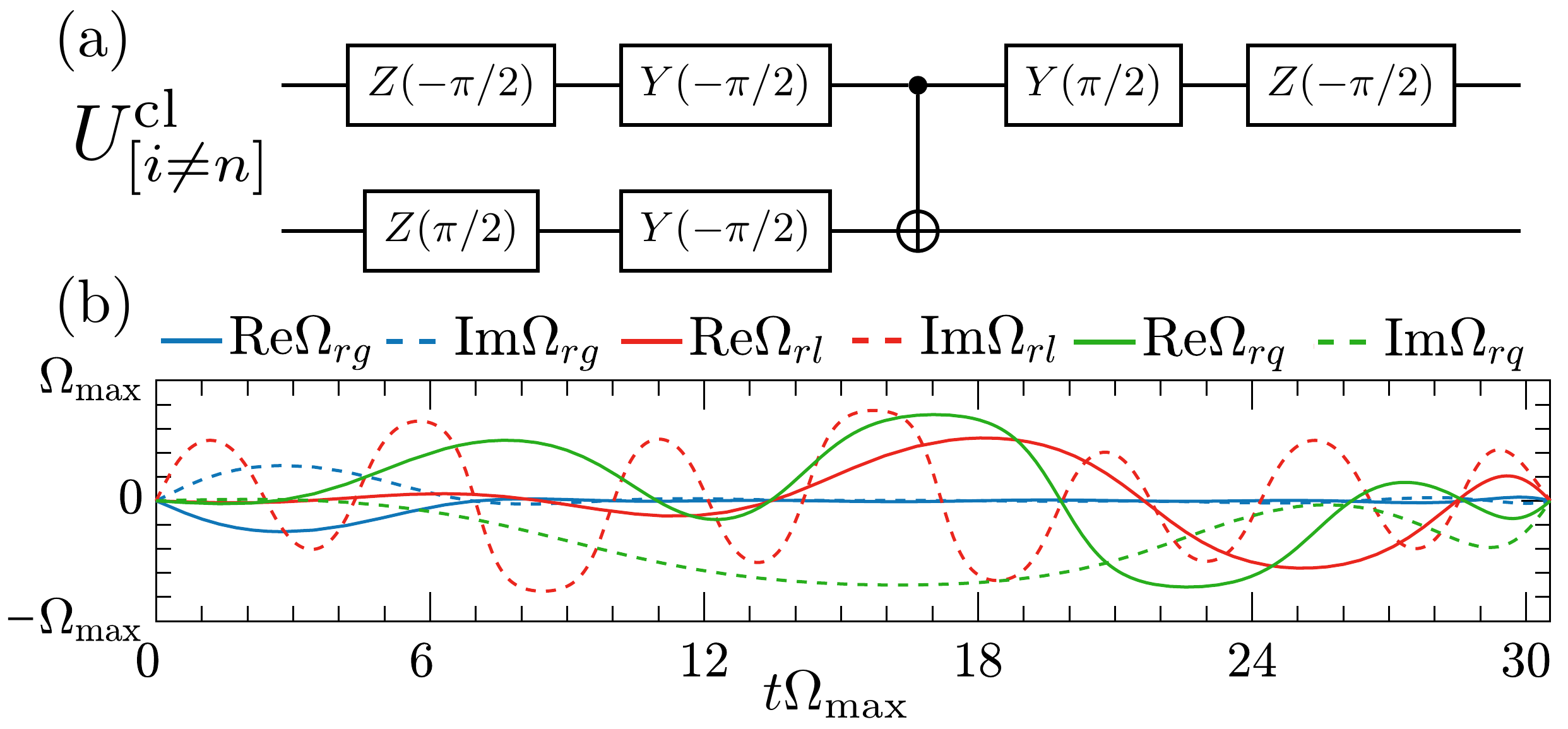}
        \caption{Implementation of the unitaries ${U^{\rm cl}_{\left[ {i \ne n} \right]}}$ required for the cluster state of \cref{cls_form} with (a) the quantum gate approach and (b) the quantum optimal control approach. The gates $X,Y,Z$ correspond to single qubit Pauli rotation, e.g. $X\left( \theta  \right) = \exp \left( {i{\sigma _x}\theta /2} \right)$. The time scale is set by the maximum Rabi frequency $\Omega_{\rm max}$ of the atomic transitions induced by the lasers $\rm{L_{q}}$ and $\rm{L_{l}}$.}
        \label{cls_circ}
\end{figure}

For example, to generate the cluster state of the form \cref{mps} with elements
\begin{equation} \label{cls_form}
\begin{gathered}
  {V^0} = \frac{1}{{\sqrt 2 }}\left( {\begin{array}{*{20}{c}}
  1&0 \\ 
  1&0 
\end{array}} \right),\quad {V^1} = \frac{1}{{\sqrt 2 }}\left( {\begin{array}{*{20}{c}}
  0&1 \\ 
  0&{ - 1} 
\end{array}} \right),\quad  \hfill \\
  \left| {{\varphi _I}} \right\rangle  = \frac{1}{{\sqrt 2 }}\left( {\left| 0 \right\rangle  + \left| {{1_q}} \right\rangle } \right),\quad \left| {{\varphi _F}} \right\rangle  = \left| 0 \right\rangle , \hfill \\ 
\end{gathered}
\end{equation}
we need to implement two kinds of unitaries ${U^{\rm cl}_{\left[ {i \ne n} \right]}}$ and ${U^{\rm cl}_{\left[ {n} \right]}}$ (see \cref{circ_unit} for details). Each application of ${U^{\rm cl}_{\left[ {i \ne n} \right]}}$ followed by $M_P$ adds one site to the cluster state, and the last unitary ${U^{\rm cl}_{\left[ {n} \right]}}$ followed by $M_P$ emits the last photon and disentangles the source from the photonic MPS. The quantum circuit for implementing ${U^{\rm cl}_{\left[ {i \ne n} \right]}}$ is shown in \cref{cls_circ}(a). ${U^{\rm cl}_{\left[ {n} \right]}}$ can be simply implemented with a two-photon Raman $\pi$-pulse that swaps all population from $\left| q \right\rangle $ to $\left| l \right\rangle $, which we discuss in detail in \cref{rb_imple}.

\subsection{Quantum Optimal Control (QOC) approach}
\label{opt_main}
The quantum gate approach can be implemented in a relatively easy fashion since one only needs to sequentially implement the gates. However, the phase gate ${S_T}\left( \theta  \right)$ is rather slow to implement, and it is less obvious how to extend this approach to universally control $\mathcal{H}_{\rm col}$ for the generation of photonic MPS with a larger bond and physical dimension $D$ and $d$. QOC allows one to implement unitary operations in high-dimensional Hilbert spaces with a speed that potentially reaches the maximum speed allowed by quantum evolution~\cite{Caneva2009, Deffner2017}, and has already been applied to Rydberg atomic array experiments~\cite{levine2019parallel,Omran570}. Thus we develop a QOC approach to implement required unitaries in our protocol based on the Gradient Optimization of Analytic Controls (GOAT) algorithm~\cite{Machnes2018}.

In our MPS generation scheme we need to construct unitaries ${U_{\left[ i \right]}}$ that behave as in \cref{u_act}. At each site $i$ of the MPS, the $D \times D$ matrices ${\{ V_{\left[ i \right]}^j\} _{j = 0,...,d - 1}}$ together form an isometry ${\hat V_{\left[ i \right]}}:\mathcal{H}_D \to {\cal H}_{{\rm{src}}}^{\left( {D,d} \right)}$ that satisfies the isometry condition $\hat V_{\left[ i \right]}^\dag {\hat V_{\left[ i \right]}} = I_D$. ${\hat V_{\left[ i \right]}}$ can be implemented through any unitary of the form
\begin{equation} \label{utarg}
{U_{\left[ i \right]}} = \left( {\begin{array}{*{20}{c}}
{{{\hat V}_{\left[ i \right]}}}&{{B_1}}\\
O&{{B_2}}
\end{array}} \right),
\end{equation}
where the first row corresponds to ${\cal H}_{{\rm{src}}}^{\left( {D,d} \right)}$ and the second to the rest of $\mathcal{H}_{\rm col}$. Here, $O$ is a zero matrix, which physically means that ${U_{\left[ i \right]}}$ does not cause the population to leak out of ${\cal H}_{{\rm{src}}}^{\left( {D,d} \right)}$. The parts $B_1$ and $B_2$ can be arbitrary as long as ${U_{\left[ i \right]}}$ is unitary. This freedom of choice pose a challenge for certain gradient-based QOC algorithms like Krotov~\cite{Palao2002} and GRAPE~\cite{Khaneja2005}, which requires backward evolution from ${U_{\left[ i \right]}}$. Thus we develop a QOC approach to efficiently synthesize ${U_{\left[ i \right]}}$ by modifying the cost function of the GOAT algorithm such that it exactly captures the $\hat V_{\left[ i \right]}^j$ and $O$ parts of $U_{\left[ i \right]}$ (see the \cref{goat_imple} for details). With this algorithm and the control Hamiltonian $H_{\rm col}\left( t \right)$, we can obtain the pulse sequence for efficiently implementing desired ${U_{\left[ i \right]}}$. As a demonstration, the pulse sequence for $U_{\left[ i \ne n \right]}^{\rm cl}$ is shown in \cref{cls_circ}(b). Here we only use the resonant laser couplings for $\mathrm{L_{g}},\mathrm{L_{q}}$ and $\mathrm{L_{l}}$, which simplifies experimental realization.

\paragraph*{Generation of high-dimensioanl MPS}
We further demonstrate the ability of generating MPS of high bond and physical dimensions by considering the following generalized GHZ state~\cite{ContrerasTejada2019a}
\begin{equation} \label{}
|\mathrm{GHZ}(n, d)\rangle=\dfrac{1}{\sqrt{d}} \sum_{i=0}^{d-1}|i\rangle_{\rm ph}^{\otimes n}.	
\end{equation}
This state serves as a quantum resource state in the resource theory of genuinely multipartite entanglement under biseparability-preserving operations~\cite{ContrerasTejada2019a}. Defining $E\left( {d,i} \right)$ as the $d$-dimensional matrix with everywhere zero but only the $i$-th diagonal element being one, $|\mathrm{GHZ}(n, d)\rangle$ can be written as an MPS of form \cref{mps} with both bond and physical dimension $D=d$ as
\begin{equation} \label{}
\begin{array}{*{20}{l}}
{V_{\left[ i \right]}^j = {V^j} = E\left( {d,j + 1} \right),\quad j \in \left( {0,...,d - 1} \right),}\\
{\left| {{\varphi _I}} \right\rangle  = \frac{1}{{\sqrt d }}\sum\limits_{i = 0}^{d - 1} {\left| {{i_q}} \right\rangle } ,\quad \left| {{\varphi _F}} \right\rangle  = \left| 0 \right\rangle .}
\end{array}
\end{equation}
This state can be sequentially generated with the following two unitaries in \cref{sysph_state}
\begin{equation} \label{u_gghz}
\begin{array}{l}
U_{\left[ {i \ne n} \right]}^{\rm gG}:\left| {{j_q}} \right\rangle  \to \left| {{j_q}{j_l}} \right\rangle ,\\
U_{\left[ n \right]}^{\rm gG}:\left| {{j_q}} \right\rangle  \to \left| {{j_l}} \right\rangle ,\quad \forall j \in \left( {0,...,d - 1} \right).
\end{array}
\end{equation}
Similar as in the cluster state case, each application of $U_{\left[ {i \ne n} \right]}^{\rm gG}$ followed by $M_P$ adds one site to the state, and $U_{\left[ {n} \right]}^{\rm gG}$ followed by $M_P$ produces the last photon and disentangles the source from the photonic MPS. The QOC pulse sequence for generating $U_{\left[ {i \ne n} \right]}^{\rm gG}$ in the case of $d=3$ is shown in \cref{goat_imple}, and $U_{\left[ {n} \right]}^{\rm gG}$ is implemented in the same way as $U_{\left[ {n} \right]}^{\rm cl}$ by a two-photon Raman $\pi$-pulse between $\left| q \right\rangle $ and $\left| l \right\rangle $. The ability of generating $|\mathrm{GHZ}(n, d)\rangle$ clearly demonstrates that our scheme is particularly suitable to create high-dimensional entanglement.

\subsection{Realistic modeling including imperfections}
\label{real_model}

There will be various imperfections in real implementations of the above photon generation protocol. During unitary operations, the dominant imperfections are related to the Rydberg state $\left| r \right\rangle $. For example, the Rydberg decay ($\Gamma_r$) and dephasing  ($\Gamma _\phi$) lead to a typical lifetime of Rydberg states around $\sim 50\mu s$~\cite{Levine2018}, which is significantly shorter than the ground-state spin-wave coherence time that already can reach sub-second level~\cite{Yang2016}. Also, the finite Rydberg non-linearity $U$ provided by the van der Waals interaction~\cite{Saffman2010} leads to a small probability to create double Rydberg excitations. The photon emission also has a finite retrieval efficiency $p_{\rm em}$. Due to the above imperfections, operating our scheme will generate a $n$-photon density matrix $\rho_{\rm ph}$ with a non-unit fidelity with respect to the targeted state ${{\cal F}_{{\rm{\rm ph}}}} = \left\langle {{\psi _{{\rm{MPS}}}}} \right|{\rho _{{\rm{\rm ph}}}}\left| {{\psi _{{\rm{MPS}}}}} \right\rangle $. Due to the sequential nature of our scheme, ${{\cal F}_{\rm ph}}$ decays exponentially with the number of photonic qudits generated $n_{\rm ph}$ as ${{\cal F}_{\rm ph}} = {e^{ - \xi  \cdot {n_{\rm ph}}}}$. Here $\xi$ represents the error per photon.

Let us denote the maximal Rabi frequency during unitary implementations in our scheme as ${\Omega _{{\rm{max}}}}$. We assume our scheme runs in the strong driving regime ${\Omega _{{\rm{max}}}} \gg {\Gamma _r}({\Gamma _\phi })$ and good blockade $U \gg {\Omega _{{\rm{max}}}}$, which has already been reached by current experiments (for example see Refs.~\cite{Levine2018,Omran570}). Including Rydberg imperfections into the modeling, we find that the system density matrix $\rho \left( t \right)$ evolves as
\begin{equation} \label{ext_me}
\begin{array}{*{20}{l}}
  \begin{gathered}
  \dfrac{{d\rho \left( t \right)}}{{dt}} =  - i\left[ {H' \left( t \right),\rho \left( t \right)} \right] \hfill \\
   + \sum\limits_{i,\alpha  = \left\{ {g,l,q} \right\}} {\dfrac{{{\Gamma _r}}}{2}\left( {2\sigma _{\alpha r}^i\rho \sigma _{r\alpha }^i - \sigma _{rr}^i\rho  - \rho \sigma _{rr}^i} \right)}  \hfill \\ 
\end{gathered}  \\ 
  { + \sum\limits_i {\dfrac{{{\Gamma _\phi}}}{2}\left( {2\sigma _{rr}^i\rho \sigma _{rr}^i - \sigma _{rr}^i\rho  - \rho \sigma _{rr}^i} \right)} ,} 
\end{array}
\end{equation}
where the doubly excited Rydberg states are included in the Hamiltonian ${H'\left( t \right)}$. Since in our scheme we need to successively implement unitaries many times and possibly reach high excitations for high bond or physical dimension, the errors will accumulate over time and lead to involved dynamics. To track the long-time evolution of our atomic array with many atoms in a reliable way, we formulate an effective model in the collective basis that takes all these errors into account (see \cref{eff_model} for details). This effective model has a Hilbert space $\mathcal H _{\rm eff}$ that contains the Rydberg doubly excited states as well as the mixed states created by decoherence.

The finite photon retrieval efficiency $p_{\rm em}$ also reduces ${{\cal F}_{\rm ph}}$. Its effect can be described by introducing the process map ${W_P}:{\mathcal{H}_{{\text{eff}}}} \to {\mathcal{H}_{{\text{eff}}}} \otimes {\mathcal{H}_{{\text{\rm ph}}}}$, that maps the system density matrix to a joint density matrix of the system and a photonic qudit (see the construction of ${W_P}$ in \cref{multi_HP_calc}). Using ${W_P}$ and the solution of the master equation in $\mathcal H _{\rm eff}$ for the unitary operation process, we can compute the photonic state fidelity ${{\cal F}_{\rm ph}}$ with the help of the matrix product density operator (MPDO)~\cite{verstraete2004matrix} approach (see \cref{mpdo_fid} for details).
\begin{figure}[h!]
    \centering
    \includegraphics[width=0.5\textwidth]{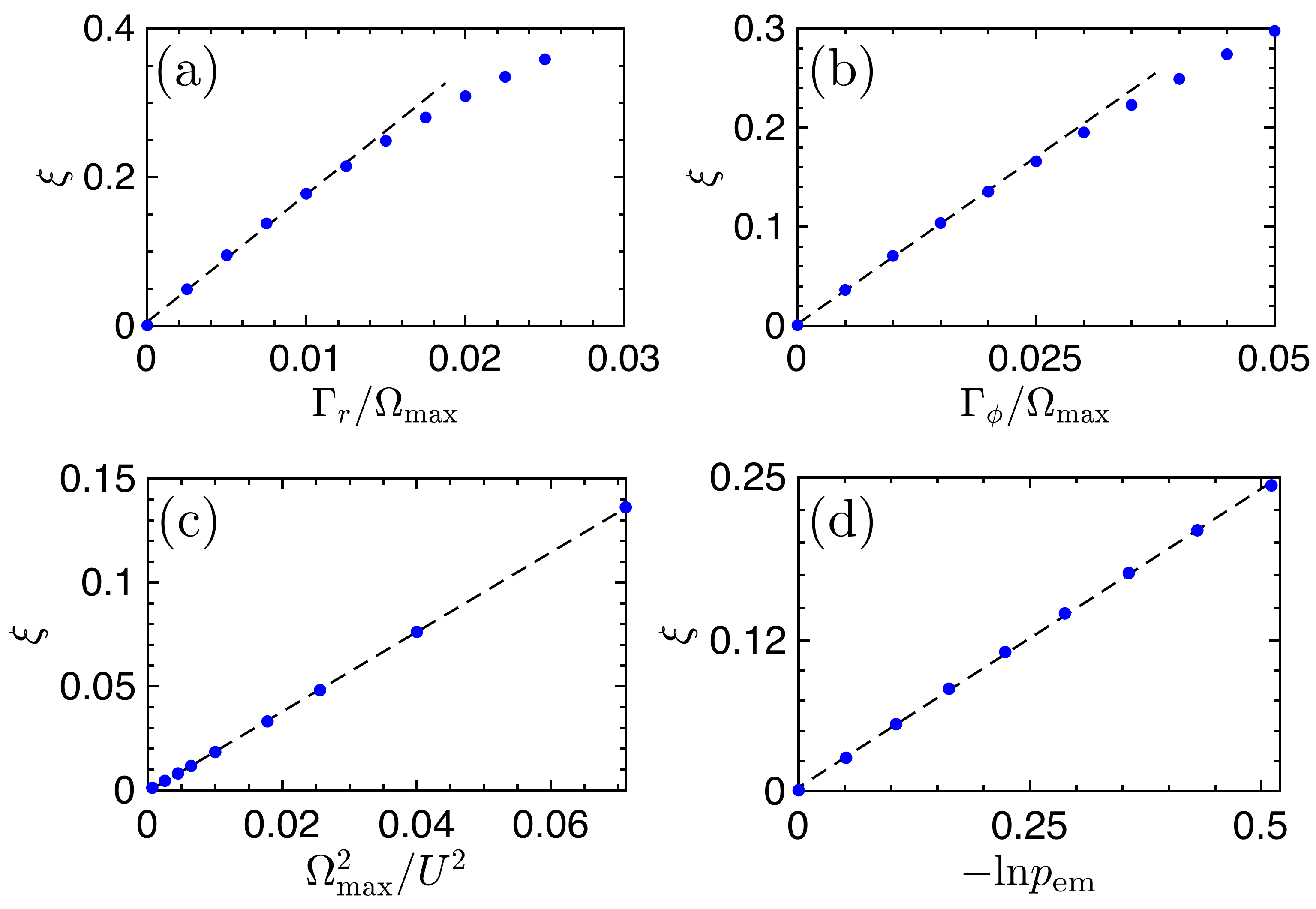}
        \caption{The dependence of $\xi $ in ${{\cal F}_{\rm ph}} = {e^{ - \xi  \cdot {n_{\rm ph}}}}$ on (a) the Rydberg decay rate ${\Gamma _r}$, (b) the Rydberg dephasing rate ${\Gamma _\phi}$, (c) the energy shift of the doubly-excited Rydberg state (denoted by $\Omega _{\max }^2/{U^2}$), and (d) the photon retrieval efficiency (denoted by $-{\rm log}p_{\rm em}$). In the strong driving regime ${\Omega _{\max }} \gg {\Gamma _r},{\Gamma _\phi }$ and with good blockade $U \gg {\Omega _{\max }}$, we obtain linear scalings.}
        \label{fid_scale}
\end{figure}

We obtain the scaling of $\xi$ by numerical simulation of the protocol to generate cluster state using the QOC pulse sequence in \cref{cls_circ}(b). By obtaining ${{\cal F}_{\rm ph}}$ as a function of photon number $n_{\rm ph}$, we can extract the error of ${{\cal F}_{\rm ph}}$ per photon $\xi  = \xi \left( {{\Gamma _r}/{\Omega _{\max }},{\Gamma _\phi}/{\Omega _{\max }},{\Omega _{\max }}/U,{p_{\rm em}}} \right)$ as shown in \cref{fid_scale}. We see that in the strong driving regime and with good blockade we have
\begin{equation} \label{fid_form}
\xi  = {\beta _0} + \frac{{{\beta _r}{\Gamma _r} + {\beta _\phi }{\Gamma _\phi }}}{{{\Omega _{\max }}}} + \frac{{{\beta _U} \cdot \Omega _{\max }^2}}{{{U^2}}} - {\beta _{{\text{em}}}} \cdot \log {p_{{\text{em}}}}.
\end{equation}
The non-universal coefficients $\left\{ {{\beta _i}} \right\}$ depend on the desired photonic state and the pulse shape. For example in \cref{fid_scale}, ${\beta _0} \approx 5 \times {10^{ - 4}}$ represents the error due to the pulse imperfection,  $\beta _r  \approx 17.1$ , $\beta _\phi  \approx 6.7$ and $\beta _U  \approx 1.9$  correspond to the Rydberg decay, dephasing and double excitation error respectively, and $\beta _{\rm em}  \approx 0.48$ corresponds to the photon retrieval error. Note that $\beta _r$ and $\beta _\phi $ can be further reduced by using a shorter pulse, at the expense of increasing $\beta _U$. In the rest of this manuscript we assume our protocol operates in the regime of small error, in which \cref{fid_form} is valid. Note that if one implement unitaries via the quantum gate approach [cf. \cref{stark_circ}], the Stark pulse will further leads to unwanted small residual population transfer~\cite{Schmidt-Kaler2004, Maller2015}, which scales as $\Omega _{\rm ST}^2/2\Delta _{\rm ST}^2$ when ${\Delta _{\rm ST}}  \gg {\Omega _{\rm ST}}$. This error does not exist in \cref{fid_form} obtained from the QOC approach.

In actual experiments, the maximum Rabi frequency ${\Omega _{\max }}$ can be tuned on demand and is only bounded by the maximal laser power available, while ${\Gamma _r},{\Gamma _\phi},U$ are determined by the properties of the atomic array. Thus, by defining the total decoherence rate $\Gamma  \equiv \beta_r {\Gamma _r} + \beta_\phi {\Gamma _\phi}$ and choosing ${\Omega _{\max }} = {\left[ {\Gamma {U^2}/\left( {2{\beta _U}} \right)} \right]^{1/3}}$, $\xi$ takes the optimized value
\begin{equation} \label{fid_ph}
{\xi _{{\text{opt}}}} = {\beta _0} + 3\beta _U^{1/3}{\left( {\dfrac{\Gamma }{{2\left| U \right|}}} \right)^{2/3}} - {\beta _{{\text{em}}}} \cdot \log {p_{{\text{em}}}}.
\end{equation}
The scaling \cref{fid_form} and \cref{fid_ph} applies to atomic arrays with arbitrary geometries as long as they lie within the Rydberg blockade radius. Different array geometries will affect the Rydberg non-linearity $U$, the photon retrieval efficiency ${p_{\rm em}}$, and the Holstein-Primakoff approximation. The analysis of array geometry dependence will be the focus of the next section.

\section{Photon retrieval in atomic arrays}
\label{pho_ret_main}

As we have shown in the previous section, we need to achieve high photon retrieval efficiency $p_{\rm em}$ because $\cal F_{\rm ph}$ decreases as $p_{{\text{em}}}^{\beta_{\rm em}  \cdot {n_{{\text{\rm ph}}}}}$. Defining the photon retrieval error as ${\varepsilon _{\rm em}} = 1 - {p_{\rm em}}$, the maximal retrieval efficiency of an atomic ensemble generally scales inversely to its optical depth OD ${\varepsilon _{{\rm{em}}}}\sim{\rm{O}}{{\rm{D}}^{ - 1}}$~\cite{Gorshkov2007,Porras2008}. In recent years, however, some theoretical works have suggested that the use of atomic arrays with optimized collective excitation profiles can lead to dramatic improvements in retrieval efficiency~\cite{asenjogarcia17a, Manzoni2018,Grankin2018}. Thus, in this section, we will characterize how the photon retrieval efficiency scales for atomic arrays and the impact of imperfections during the photon emission process. See \cref{apd_retri} for more details about this section.

The photon retrieval process is shown in \cref{fig1}(a), which consists of a Raman pulse of Rabi frequency ${\Omega _{el}}\left( t \right)$ and detuning $\Delta \left( t \right)$ that transfers the population on $\left| l \right\rangle $ to a fast-decaying excited state $\left| e \right\rangle $, which only decays to the state $\left| g \right\rangle $. During the emission process $\left| e \right\rangle  \to \left| g \right\rangle $, the emitted photon can be rescattered by atoms due to the dipole-dipole interactions between atoms~\cite{Lehmberg1970}. Starting from a singly excited state 
\begin{equation} \label{em_init}
\left| {{1_l}} \right\rangle  = S_{l,{{\bf{k}}_{rg}} - {{\bf{k}}_{rl}}}^\dag \left| 0 \right\rangle  = \sum\limits_j {{u_j}{e^{i\left( {{{\bf{k}}_{rg}} - {{\bf{k}}_{rl}}} \right) \cdot {{\bf{r}}_{{j}}}}} \sigma _{lg}^j} {\left| g \right\rangle ^{ \otimes N}},
\end{equation}
the decay dynamics is governed by the following non-Hermitian Hamiltonian ${H_{\rm em}}(t) = {H_{el}}(t) + {H_{{\rm{DDI}}}}$ with
\begin{equation} \label{nonh_ham}
\begin{array}{l}
{H_{el}}\left( t \right) = \frac{{{\Omega _{el}}\left( t \right)}}{2}\sum\limits_i {\left( {\sigma _{el}^i{e^{i{{\bf{k}}_{el}}\cdot {{\bf{r}}_i}}} + {\rm H.c.}} \right)}  - \Delta (t)\sum\limits_i {\sigma _{ee}^i} ,\\
{H_{{\rm{DDI}}}} =  - {\mu _0}d_{eg}^2\omega _{eg}^2\sum\limits_{j,l} { {\bf{d}}_j^*}  \cdot {{\bf{G}}_0}\left( {{{\bf{r}}_j},{{\bf{r}}_l},{\omega _{eg}}} \right) \cdot { {\bf{d}}_l}\sigma _j^{eg}\sigma _l^{ge}.
\end{array}
\end{equation}
Here, ${d_{eg}}$ and ${ {\bf{d}}_j}$ are the dipole matrix element and unit atomic polarization vector. The dyadic Green's tensor is~\cite{novotny2012principles}
\begin{equation} \label{dya_green}
\begin{aligned}
  {{\mathbf{G}}_0}\left( {{{\mathbf{r}}_j},{{\mathbf{r}}_l},{\omega _{eg}}} \right) = \frac{{{{\text{e}}^{{\text{i}}{k_0}R}}}}{{4\pi k_0^2{R^3}}}\left[\left( {k_0^2{R^2} + {\text{i}}{k_0}R - 1} \right){\mathbf{I}} \hfill \right. \\
\left.   + \left( { - k_0^2{R^2} - 3{\text{i}}{k_0}R + 3} \right)\frac{{{\mathbf{R}} \otimes {\mathbf{R}}}}{{{R^2}}}\right],
\end{aligned}
\end{equation}
with ${\bf {R = r}}_{j} - {\bf{r}}_{l}$ and $k_0={\omega _{eg}}/c$ with $c$ being the speed of light. From \cref{nonh_ham} and \cref{dya_green} we can compute the evolution of the atomic array and the photon retrieval efficiency following the approach in Ref.~\cite{Manzoni2018}. The photon retrieval efficiency is defined as the probability of the photon to be emitted into some detection mode ${{\bf{E}}_{{\rm{det}}}}(\bf r)$.

To achieve high retrieval efficiency, one needs to optimize the excitation profile of the collective excitation to best match the detection mode~\cite{asenjogarcia17a,Manzoni2018, Grankin2018}. We study two types of excitation profiles. One is the optimal profile as studied in Ref.~\cite{Manzoni2018}, which gives the minimal photon retrieval error $\varepsilon _{\rm em}^{\rm opt}$. Another is the profile created by a Gaussian beam, which we call the Gaussian profile, with its photon retrieval error denoted as $\varepsilon _{\rm em}^{\rm Gauss}$. To assess our photonic MPS generation protocol, we only use the Gaussian profile, as it is experimentally easier to implement.

\begin{figure*}
	\centering
	\includegraphics[width=\textwidth]{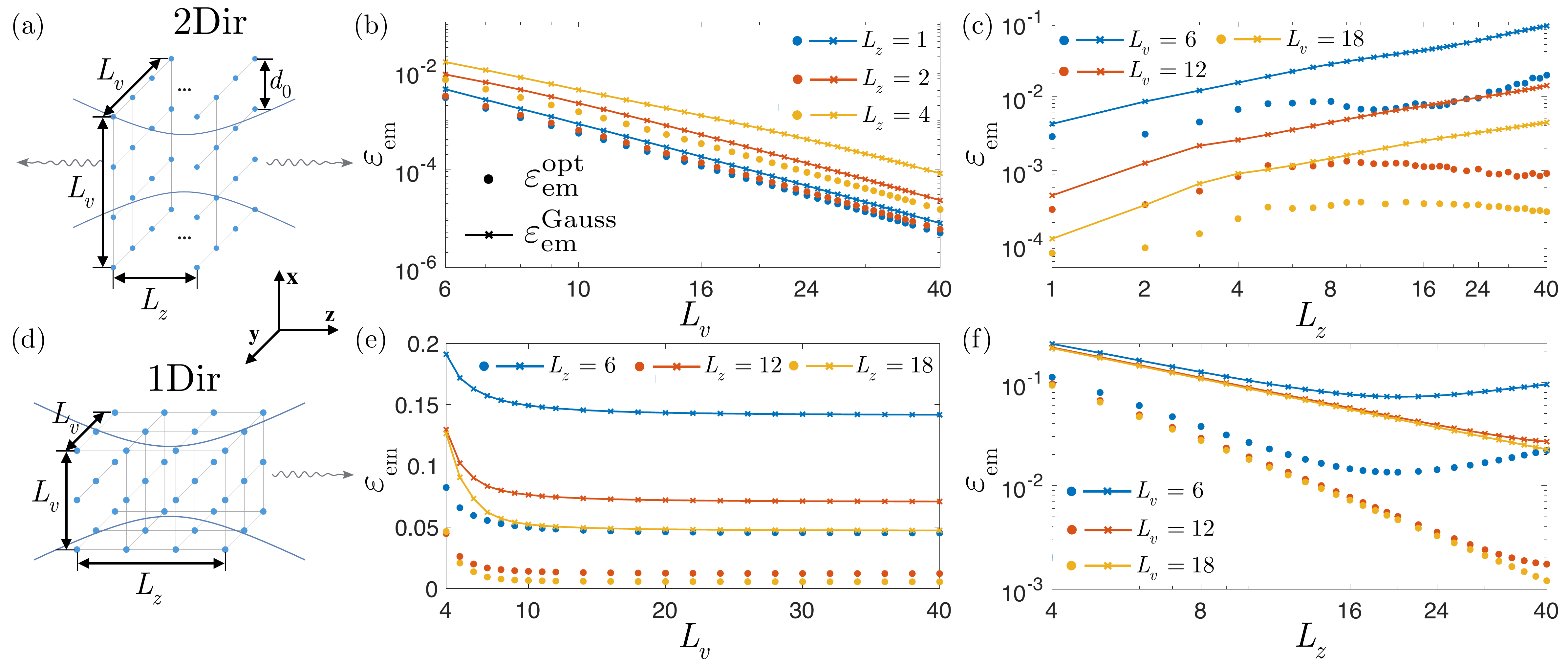}
        \caption{The scaling of the photon retrieval error from excitation with optimal profile $\varepsilon _{\rm em}^{\rm opt}$ and with Gaussian profile $\varepsilon _{\rm em}^{\rm Gauss}$ as array geometries ${L_v}$ and ${L_z}$ for an interatomic distance $d_0=0.6\lambda_{eg}$. (a) A schematic plot of two-directional retrieval. We denote the scheme by `2Dir' in the plot. (b) and (c) are the scaling of $\varepsilon _{\rm em}^{\rm opt}$ and $\varepsilon _{\rm em}^{\rm Gauss}$ with ${L_v}$ and $L_z$ for two-directional retrieval scheme. Similarly, (d),(e),(f) are the schematic plots and scaling of photon retrieval errors with ${L_v}$ and $L_z$ for the uni-directional retrieval scheme.}
        \label{retri_results}
\end{figure*}

For concreteness, we fix the distance between atoms to be $d_0 = 0.6\lambda_{eg} $, but similar subwavelength interatomic distances give similar results. We consider two types of retrieval schemes. In the scheme shown in \cref{retri_results}(a), the photon is emitted in two opposite directions, where the detection mode is a symmetric superposition of two vector Gaussian modes~\cite{Chen2002a} along the photon emission directions. This retrieval setup has been shown to lead to favorable scaling of the retrieval efficiency in the case of a single layer two-dimensional array~\cite{Manzoni2018}. We assume that the photons collected from these two directions can be later recombined, for example using the setup in~\cite{Bekenstein}. The geometry of our atomic array is characterized by its transverse size ${L_v}$ and longitudinal size ${L_z}$.

The \cref{retri_results}(b) shows the scaling of the photon retrieval error as a function of ${L_v}$ for the two-directional retrieval scheme. For the case of ${L_z} = 1$ we reproduced the scaling $\varepsilon _{\rm em}^{\rm opt} \approx {\left( {\log {L_v}} \right)^2}/L_v^4$ in Ref.~\cite{Manzoni2018}. When increasing $L_z$, we obtain $\varepsilon _{\rm em}^{\rm opt} \approx c^{\rm opt}_{{L_z}} \cdot {\left( {\log {L_v}} \right)^2}/L_v^4$. For the Gaussian excitation profile, we also have $\varepsilon _{\rm em}^{\rm Gauss} \approx c^{\rm Gauss}_{{L_z}} \cdot {\left( {\log {L_v}} \right)^2}/L_v^4$ when $L_z$ is small. This indicates that for the two-directional retrieval scheme with an array of a few layers in the transverse directions, the optimal excitation profile is close to a Gaussian profile. 

In \cref{retri_results}(c) we show the behavior of the photon retrieval error as a function of ${L_z}$. When $L_z$ increases, more of the photonic field can leak out through the transverse directions that are perpendicular to the detection mode. Thus, in general, the photon retrieval efficiencies increase when ${L_z}$ increases. For the Gaussian profile, one obtains a power-law scaling, that $c_{{L_z}}^{{\text{Gauss}}} \approx L_z^{0.7}$. In the optimal case, the behavior is more involved, since in this case the excitation profile adapts to the detection mode, which partially compensates for the leakage of the photonic field.

The other retrieval scheme we consider is a typical uni-directional retrieval shown in \cref{retri_results}(d). In this case, the detection mode is a single vector Gaussian beam. In \cref{retri_results}(e), we show the scaling of the photon retrieval errors $\varepsilon _{\rm em}^{\rm opt}$ and $\varepsilon _{\rm em}^{\rm Gauss}$ as a function of ${L_v}$. Apart from the region where ${L_v}$ is small, $\varepsilon _{\rm em}^{\rm opt}$ and $\varepsilon _{\rm em}^{\rm Gauss}$ do not depend on ${L_v}$, which is expected from the result of the atomic ensemble, since the optical depth does not involve the transverse size. In \cref{retri_results}(f) we show the scaling of $\varepsilon _{\rm em}^{\rm opt}$ and $\varepsilon _{\rm em}^{\rm Gauss}$ as a function of ${L_z}$. We get ${\varepsilon^{\rm opt} _{\rm em}} \approx {\rm{1.45}}L_z^{ - 2}$ and ${\varepsilon^{\rm Gauss} _{\rm em}} \approx {\rm{0.84}}L_z^{ - 1}$ when  ${L_v}/L_z$ is not very small. Here the scaling of  $\varepsilon _{\rm em}^{\rm Gauss}$ is already reaching the best scaling for the atomic ensemble case~\cite{Gorshkov2007,Porras2008}. We further find that in a wide range of geometries $L_v$ and $L_z$, the $\varepsilon^{\rm Gauss}_{\rm em}$ of uni-directional scheme can be related to that of the two-directional scheme through
\begin{equation} \label{Gauss_err_sum}
\varepsilon _{\rm em}^{\rm Gauss}\left( \textrm{1Dir} \right) \approx 0.84{L^{-1}_z} + \varepsilon _{\rm em}^{\rm Gauss}\left( \textrm{2Dir} \right),
\end{equation}
which indicates that the error ${\varepsilon_{\rm em}^{\rm Gauss}}$ of the uni-directional retrieval scheme stems from both the photonic field leaked from the direction opposite to the detection mode, which scales as $0.84{L^{-1}_z}$, and the field leakage to the transverse directions, which share the same scaling as the error ${\varepsilon_{\rm em}^{\rm Gauss}}$ for the two-directional retrieval scheme. This behavior predicts an increase of $\varepsilon _{\rm em}^{\rm Gauss}$ when ${L_v}/{L_z}\ll 1$, where the leakage to transverse direction is the dominant error. Such behavior is observed in \cref{retri_results}(f) for $L_v=6$ case, and qualitatively explains the increase of $\varepsilon _{\rm em}^{\rm opt}$ in that region as well.

The scaling $\varepsilon _{\rm em}^{\rm Gauss} \propto L_z^{ - 1}$ of the uni-directional retrieval error with Gaussian profile indicates that it shares the same physics as the photon retrieval process in disordered atomic ensembles where the optical depth ${\rm OD} \propto {L_z}$.
Thus, just like the atomic ensemble case, we can add a cavity to enhance the optical depth of the atomic array. Assuming we add a cavity of finesse $F$, the photon retrieval error is reduced approximated $F$-fold. Such a cavity-enhanced photon retrieval from an optical lattice has been demonstrated in Ref.~\cite{Bao2012,Yang2016} with $F \approx 50$.

Imperfections during the photon emission process include (i) array defects, (ii) thermal atomic motion, and (iii) the deviation of the Holstein-Primakoff approximation. Here we briefly describe their effects, and the details are shown in \cref{multi_HP_calc} and in \cref{apd_retri}.

\paragraph{Atomic defects.}
As shown in Ref.~\cite{Manzoni2018}, one expects that the relative decrease of the retrieval efficiency should be proportional to the ratio of the detection mode hitting the empty sites. Numerical results show that the following relation for the retrieval efficiency $p_{\rm em}^{\rm def}$ with defects holds \cite{Manzoni2018}
\begin{equation} \label{p_def}
p_{\rm em}^{\rm def} = {p_{\rm em}}\left( {1 - \alpha_{\rm def} \frac{{\sum\limits_{j \in {\rm{def}}} {{{\left| {{E_j}} \right|}^2}} }}{{\sum\limits_l {{{\left| {{E_l}} \right|}^2}} }}} \right),
\end{equation}
where ${E_j} = {{\bf{E}}_{{\rm{det}}}}\left( {{{\bf{r}}_j}} \right) \cdot {\bf{d}}_j^*$ is the intensity of the detection mode at the position of the $j$-th atom, and the coefficient $\alpha_{\rm def}$ depends on $d_0$ and the array geometry. 
\paragraph{Atomic thermal motion.}
The effect of atomic thermal motion is modeled as a random spatial disorder with a standard deviation of $\sigma_{\rm th}$ \cite{Duan2002}.
This leads to an additional photon retrieval error that scales as $\sigma _{{\rm{th}}}^2/{d_0^2}$, the same as predicted in Refs.~\cite{Manzoni2018, Shahmoon2017}.

\paragraph{Holstein-Primakoff approximation ~\cite{Holstein1940}.}
The error due to the approximation [\cref{ops_approxi}] comes in when our photonic MPS generation protocol in \cref{seq_gen} involves states with multiple collective excitations. This error lead to a lower bound on the photon retrieval efficiency for generating MPS with bond dimension $D$ and physical dimension $d$ in the regime of $D + d -2 \ll N$ as
\begin{equation} \label{reno_pem}
{p_{{\rm{em}}}} \to {p'_{{\rm{em}}}} \ge {\left( {1 - \frac{{\left( {D + d - 2} \right)\left( {D + d - 3} \right)}}{2N}} \right)^{1/\left( {d - 1} \right)}}{p_{{\rm{em}}}}.
\end{equation}
 For the cluster state, we have $D=2$ and $d=2$, thus ${p'_{{\rm{em}}}} \ge \left( {1 - \frac{1}{N}} \right){p_{{\rm{em}}}}$.

\section{Optimal performance of the scheme}
\label{tot_pho_opt}

Given that the error per photon ${{\xi _{{\rm{opt}}}}}$ scales as \cref{fid_ph} and with the retrieval efficiencies studied in \cref{pho_ret_main}, we can optimize the array geometry ${L_v}$ and ${L_z}$ to find the minimal ${\xi _{{\text{opt}}}}$.

The array geometry will affect both the retrieval efficiency $p_{\rm em}$ and the Rydberg nonlinear shift $U$. The $U$ is determined by the van der Waals interaction ${V_{ij}} = {C_6}/{\left| {{{\vec r}_i} - {{\vec r}_j}} \right|^6}$ between Rydberg atoms at position ${\vec r_i} = {d_0}\vec i = {d_0}\left( {{i_x},{i_y},{i_z}} \right)$ and ${\vec r_j}$ as~\cite{Walker2008}
\begin{equation} \label{ushift1}
U = \sqrt {\frac{{N\left( {N - 1} \right)}}{{\sum\limits_{\vec i \ne \vec j} {{{\left| {\vec i - \vec j} \right|}^{12}}} }}} \frac{{{C_6}}}{{d_0^6}} \equiv {f_{{L_v},{L_z}}}\frac{{{C_6}}}{{d_0^6}},
\end{equation}
where we denote the geometric factor as ${f_{{L_v},{L_z}}}$. Substituting \cref{ushift1} and $N = L_v^2{L_z}$ into \cref{fid_ph} and using the lower bound of $p_{\rm em}$ in \cref{reno_pem}, we get
\begin{equation} \label{err_opt_mid}
{\xi _{{\rm{opt}}}}{\rm{ }} = {\beta _0} + {\left( {\frac{{27{\beta _U}}}{{4f_{{L_v},{L_z}}^2}}} \right)^{1/3}}{\left( {\frac{{\Gamma d_0^6}}{{\left| {{C_6}} \right|}}} \right)^{2/3}}{\rm{ }} - {\beta _{{\rm{em}}}} \cdot \log {{p'_{{\rm{em}}}}}.
\end{equation}

Since in experiments the parameters ${\Gamma},{C_6},{d_0}$ depend on the particular atomic configuration chosen, in our numerical optimization we scan over $L_v$ and $L_z$ to find the minimal $\xi_{\rm opt}$ with respect to a given $\left| {{C_6}} \right|/\left( {\Gamma d_0^6} \right)$. To quantify the performance more intuitively, we define an \textit{entanglement length} ${N_{\rm ph}} \equiv \log 2/\xi_{\rm opt}$, which is the photon number that can be generated with ${\cal F}_{\rm ph}=1/2$.

\begin{figure}[h!]
	\centering
	\includegraphics[width=0.49\textwidth]{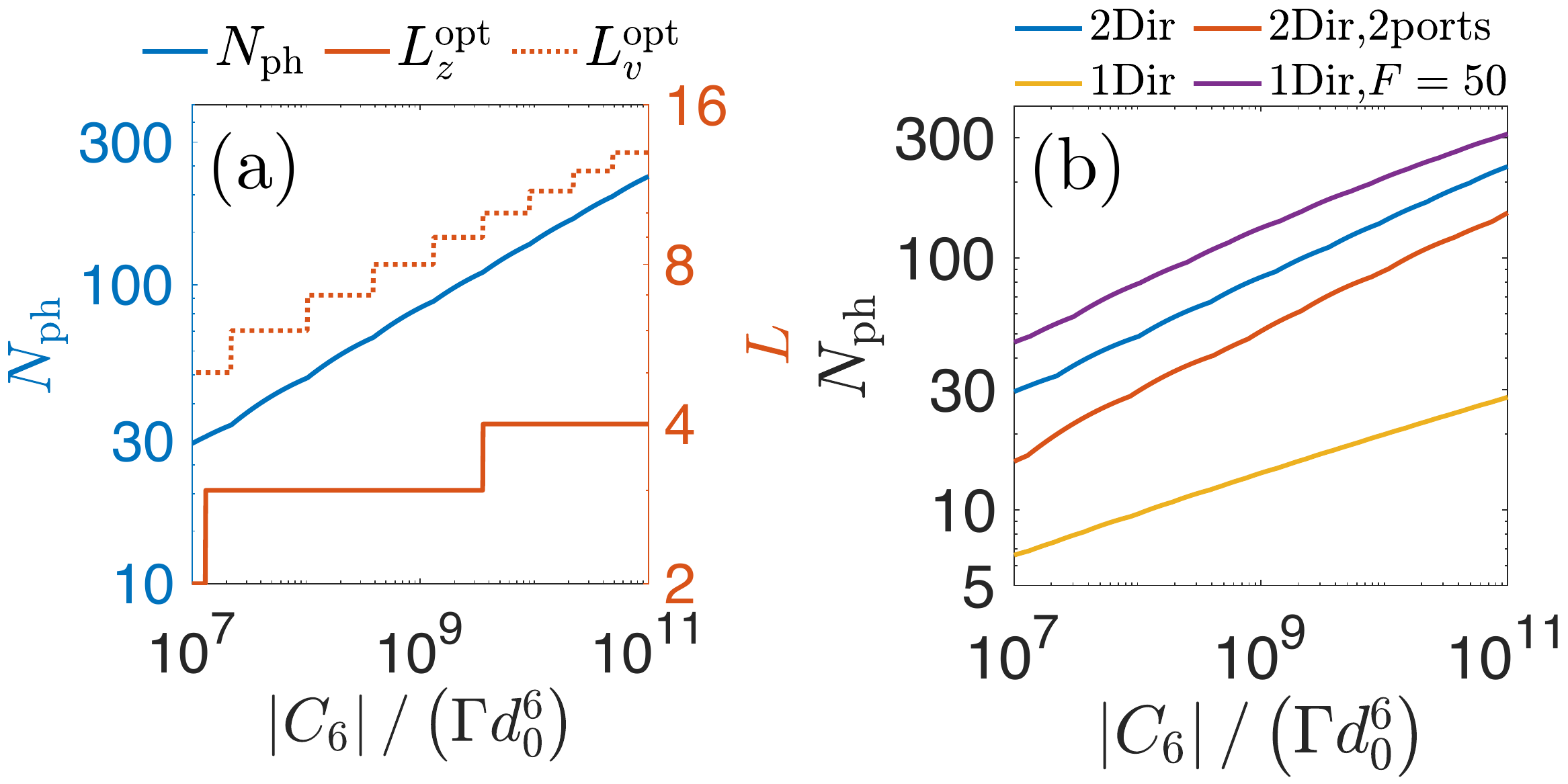}
        \caption{(a) The scaling of the entanglement length $N_{\rm ph}$ and corresponding optimal array geometry $L^{\rm opt}_v$ and $L^{\rm opt}_z$ as a function of overall resource $\left| {{C_6}} \right|/\Gamma d_0^6$ for the two-directional photon retrieval scheme. (b) A comparison of the $N_{\rm ph}$ scaling for different photon retrieval schemes, including the uni-directional and two-directional retrieval schemes in free space, the one-directional retrieval scheme assisted by a cavity with finesse $F=50$, and the two-port two-directional retrieval scheme [cf.~\cref{fse}]. The current state-of-the-art experimental parameters [cf.~\cref{exp_para}] gives $\left| {{C_6}} \right|/\left( {\Gamma d_0^6} \right) = 7.9 \times {10^7}$.}
        \label{opt_scaling_a}
\end{figure}

As a demonstration, the result of this optimization for generating cluster state with optimal control pulse in \cref{cls_circ}(b) is shown in \cref{opt_scaling_a}. For the two-directional retrieval scheme, we obtain ${N_{{\rm{\rm ph}}}}\sim{[ {| {{C_6}} |/( {\Gamma d_0^6} )} ]^{0.23}}$, shown in \cref{opt_scaling_a}(a). The corresponding optimal array geometry $L_z^{\rm opt}$ and $L_v^{\rm opt}$ can also be estimated from a power law scaling $L_z^{\rm opt} \sim{[ {| {{C_6}} |/( {\Gamma d_0^6} )} ]^{0.05}}$ and $L_v^{\rm opt} \sim{[ {| {{C_6}} |/( {\Gamma d_0^6} )} ]^{0.1}}$. We find such power law scaling behavior is universal for both uni-directional and two-directional retrieval schemes, with different exponents depending on the non-universal constants $\{ {{\beta _i}} \}$ and the specific retrieval scheme. We further compare the performance of different schemes in \cref{opt_scaling_a}(b), from which we see the uni-directional retrieval + cavity scheme and two-directional retrieval scheme gives rise to favourable scaling of $N_{\rm ph}$. To make a quantitative estimation of $N_{\rm ph}$, we use current state-of-the-art experimental parameters~\cite{Levine2018}, in which the Rydberg state is $| {70S,J = 1/2,{m_J} =  - 1/2} \rangle $, with
\begin{equation} \label{exp_para}
\begin{gathered}
  {C_6} = { - 862}.{\text{69GHz}} \cdot \mu {{\text{m}}^{\text{6}}}, \hfill \\
  \Gamma _{r}=19.6\text{kHz},\quad {\Gamma _\phi}=21.3\text{kHz}, \hfill \\ 
\end{gathered}
\end{equation}
and assume we work on a optical lattice with ${d_0} =532\rm{nm}$~\cite{Zeiher2015}, which leads to $\left| {{C_6}} \right|/\left( {\Gamma d_0^6} \right) = 7.9 \times {10^7}$. With this parameter set, the uni-directional protocol in free space can generate a cluster state of ${N_{\rm ph}} \approx 9$ photons with fidelity ${{\cal F}_{\rm ph}}=1/2$. If we add a ring cavity of finesse $F=50$ (similar to that used in Ref.~\cite{Yang2016}), this improves to ${N_{\rm ph}} \approx 75$ photons. While for the two-directional retrieval scheme, this parameter set yields ${N_{\rm ph}} \approx 47$ photons. We estimate the required maximal Rabi frequency ${\Omega _{\max }}\sim 2\pi  \times \rm{15MHz}$ for the this set of parameters, already available in current experiments~\cite{Madjarov2020a}. Depending on the available laser power, one could synthesize pulses that lead to smaller values of ${\beta _r}( {{\beta _\phi }} )/{\beta _U}$ compared to that in \cref{cls_circ}(b) to reduce ${\Omega _{\max }}$.

These results indicate that using the uni-directional retrieval + cavity scheme or the two-directional retrieval scheme, our scheme can deterministically generate strongly-entangled photonic MPS with a large number of photons. We further qualitatively estimate that ${N_{\rm ph}}\sim{D^{ - 2}}{d^{ - 2}}$ in \cref{dim_scale}. The entanglement length $N_{\rm ph}$ can be improved by reducing $\Gamma d_0^6/\left| {{C_6}} \right|$, which can be done by choosing a Rydberg state with a larger principal number, making the interatomic distance $d_0$ smaller, or reducing the decoherences of the Rydberg excitation. Using a cavity with larger finesse $F$ also improves $N_{\rm ph}$. We also point out that our analysis based on the atomic array can be applied to the case of a disordered atomic ensemble as well.

\section{Free-space Multi-port Emission}
\label{fse}
An important ingredient of the quantum network is to create large-scale distributed quantum entanglement. In principle, thanks to the sequential nature of our protocol, we can generate temporally separated photons and send them to a photon switch within an optical fiber to distribute photons into different ports within the fiber. Here we go one step further to propose a setup and protocol that can directly distribute the sequentially generated photons into multi-ports in free space.

\begin{figure}[h!]

	\centering
	\includegraphics[width=0.49\textwidth]{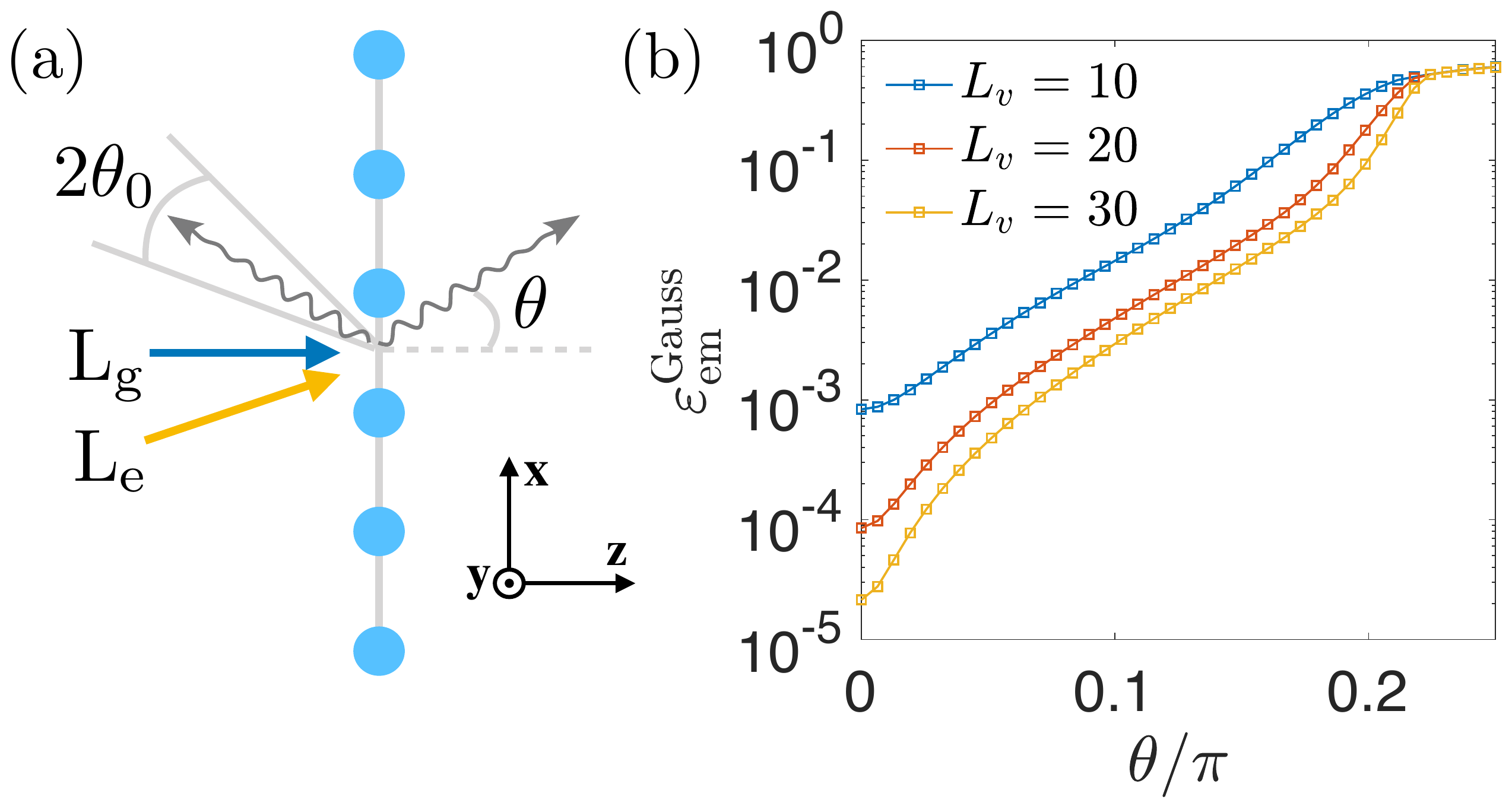}
        \caption{A multi-port free-space multiphoton source. (a) The schematic plot of the multi-port retrieval scheme. The laser $\rm {L_g}$ is perpendicular to the plane to create collective Rydberg excitation, and the laser $\rm {L_e}$ can be shone from multiple directions to control the direction of the emitted photon(s). Here for clarity we did not plot the lasers $\rm {L_l}$ and $\rm {L_q}$. (b) The angle dependence of the photon retrieval error $\varepsilon _{\rm em}^{\rm Gauss}$ of the excitation with Gaussian profile.}
        \label{angle_result}
\end{figure}

In our MPS generation protocol, the direction of photon emission can be controlled by the photon retrieval laser $\rm L_e$, which imprints a plane-wave phase pattern with a specific momentum. Due to the conservation of momentum, the emitted photon tends to fly to a phase-matched direction~\cite{Porras2008}. Based on this effect, we propose the following setup depicted in \cref{angle_result}(a) for the multi-port photon generation, where we have a one-layer, two-dimensional square array. The laser $\rm{L}_g$ is perpendicular to the plane to create collective Rydberg excitation, and the lasers $\rm{L}_q$ ($\rm{L}_l$) with plane-wave profile with fixed momenta is used to transfer the collective exciation between $\left| r \right\rangle $ and $\left| q \right\rangle $ ($\left| l \right\rangle $). The emission direction of photons is determined by the momentum of the laser $\rm L_e$. The photon(s) will fly to two directions that have the same momentum component parallel to the plane, and opposite momentum component perpendicular to the plane, characterized by the angle $\theta$. Thus we choose the detection mode to be a symmetric superposition of two vector Gaussian modes along the photon emission directions.

The photon retrieval error of the excitation with Gaussain profile $\varepsilon _{{\rm{em}}}^{{\rm{Gauss}}}$ as a function of $\theta$ is characterized in \cref{angle_result}(b). As expected, $\varepsilon _{{\rm{em}}}^{{\rm{Gauss}}}$ gradually increases with larger angle $\theta$. The emitted photonic field along each direction is well characterized by a Gaussian beam~\cite{Porras2008,Manzoni2018} with wavelength $\lambda_{eg}$, beam waist $w_0$, and asymptotic beam angle $\theta_0  = \lambda_{eg}/ {\pi {w_0}}$. In order to make the photons in different ports distinguishable, the angular distance between different ports should at least be $2{\theta _0}$. The minimal version of the multi-port device consist of two ports with angles ${\theta _1} = {\theta _0}$ and ${\theta _2} = -{\theta _0}$. Thus we substitute the retrieval efficiency $p_{\rm em}$ at the angle $\theta  = {\theta _0}$ into \cref{err_opt_mid} and optimize $L_v$ to get the maximal entanglement length $N_{\rm ph}$. The scaling of $N_{\rm ph}$ for the two-port device is shown in \cref{opt_scaling_a}(b), and we get ${N_{{\rm{\rm ph}}}}\sim{[ {| {{C_6}} |/( {\Gamma d_0^6} )} ]^{0.24}}$ for this case, with corresponding optimal array size $L_v^{{\rm{opt}}}\sim{[ {| {{C_6}} |/( {\Gamma d_0^6} )} ]^{ - 0.1}}$ (data not shown). With the current experimental parameters [\cref{exp_para}], we can generate cluster state of $N_{\rm ph}\approx 28$ with this two-port device. By allowing photon emission with larger angle $\theta$, one can increase the number of ports available at the expense of increasing photon retrieval error. We also point out that such multi-port photon retrieval works for three-dimensional atomic arrays as well \cite{Porras2008}, where the emission direction is further affected by the Bragg scattering along the $z$ direction in \cref{angle_result}(a).

\section{Experimental Consideration}
\label{rb_imple}
\begin{figure}[h!]
    \centering
    \includegraphics[width=0.4\textwidth]{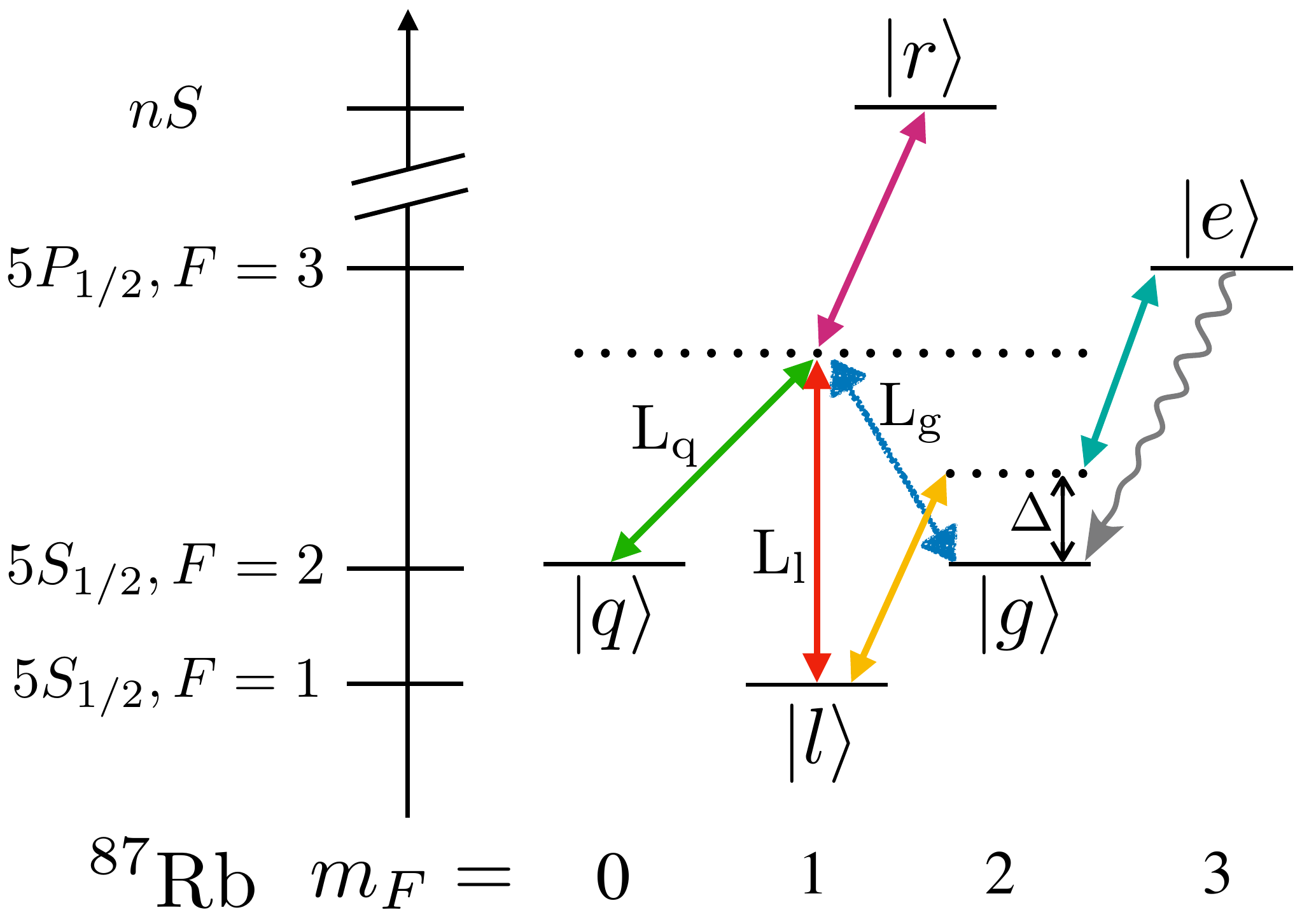}
        \caption{An implementation of the level scheme in \cref{fig1}(a) using $^{87}\textrm{Rb}$ level structure. The hyperfine ground states are individually coupled to the Rydberg state $\left| r \right\rangle $ via two-photon transitions with lasers of different frequencies and polarizations. The on-demand photon emission is triggered by a two-photon Raman transition from $\left| l \right\rangle$ to $\left| e \right\rangle $ \cite{Porras2008}.}
        \label{rb_pol}
\end{figure}

The level scheme in \cref{fig1}(a) can be realized in various types of atoms. In \cref{rb_pol} we illustrate a possible level scheme based on rubidium-87 ($^{87}\textrm{Rb}$), with quantum number of relevant states marked on the figure. The Rydberg state could be $\left| r \right\rangle  = \left| {70S,J = 1/2,{m_J} =  1/2} \right\rangle $, for example.

The experiment can be initialized by preparing all atoms in the state ${\left| g \right\rangle ^{ \otimes N}}$ through optical pumping. Each state can be coupled to the Rydberg state with two-photon transitions (shown as arrows with different colors in \cref{rb_pol}). By choosing the frequencies and the polarizations of different lasers, one can address the transitions depicted in \cref{rb_pol} individually. As a particular example, one can directly couple the state $\left| l \right\rangle $ and $\left| q \right\rangle $ by two-photon Raman transition using the lasers $\rm{L_l}$ and $\rm{L_q}$ shown in \cref{rb_pol}. This allows us to simply implement the $U_{\left[ n \right]}^{\rm cl}$ and $U_{\left[ n \right]}^{\rm gG}$ discussed in \cref{unit_imple}.

In \cref{rb_pol}, the on-demand photon emission is realized by a two-photon Raman transition that couples $\left| {{l}} \right\rangle  \to \left| {{e}} \right\rangle $ through the intermediate state $\left| {{g}} \right\rangle$ with a detuning ${\Delta}$~\cite{Porras2008}. Due to the selection rules on $m_F$, the photon decays from $\left| {{e}} \right\rangle$ only to $\left| {{g}} \right\rangle$. In the case of generating photonic MPS with physical dimension $d=2$, at most a single photon is emitted each time. In that case, one can alternatively use a laser to couple state $\left| r \right\rangle $ and $\left| e \right\rangle$, go through the process $\left| l \right\rangle  \to \left| r \right\rangle  \to \left| e \right\rangle $ to emit photons. In this case it is further possible to operate the protocol without using the state $\left| l \right\rangle $, which could simplify the experimental realization, at the expense of storing the collective excitation on $\left| r \right\rangle $ that has a larger decoherence rate.

\section{Summary and Outlook}
\label{sum_olk}

Summing up, we have proposed a physical platform and a protocol to deterministically generate multiphoton entangled states in free space based on a Rydberg-blockaded atomic array. We have developed both a quantum gate approach and the quantum optimal control approach to universally control a large Hilbert space, which enables the generation of photonic MPS with large bond and physical dimensions, and allows one to experimentally explore the rich regime of one-dimensional spin systems. The strong nonlinearity and long coherence time of the Rydberg state together with the good photon retrieval efficiencies of atomic arrays, lead to favorable scaling of the entanglement length of the photonic MPS, which may enable the generation of large-scale photonic quantum entanglement beyond the state of the art. Furthermore, one can control the photon emission angle of the sequentially generated photons, which makes this platform a potential multi-port device that can efficiently distribute entangled photons in free-space.

This work can be extended in many ways. First, one can further reduce the decoherence errors by using long-lived states in alkaline earth atoms~\cite{Dunning2016}, optimize the pulse shape, and apply open system optimal control techniques~\cite{Schulte-Herbruggen2011,Machnes2018,Abdelhafez2019}. Second, a more involved atomic level scheme may allow one to produce photonic MPS with polarization encoding~\cite{Porras2008,zyth2019}, and the photon directional retrieval enables momentum multiplexing~\cite{Cox2019,Porras2008}, which can be used to generate higher-dimensional photonic tensor network states in free space~\cite{zyth2019}. The ability to generate tensor network states with high bond dimension and physical dimension further call for identifying genuine quantum resource states in this class of states that can be efficiently created in the near-term devices. Finally, the principle to obtain a large Hilbert space with bosonic modes and the associated control techniques developed here can be extended to various other platforms, for example waveguide QED~\cite{Gonzalez-Tudela2015} and circuit QED systems~\cite{Fink2008a}.

\begin{acknowledgments}
We thank Jun Rui and Tao Shi for insightful discussions. Z.-Y.W, D.M., and J.I.C. acknowledge funding from ERC Advanced Grant QENOCOBA under the EU Horizon 2020 program (Grant Agreement No. 742102) and within the D-A-CH Lead-Agency Agreement through project No. 414325145 (BEYOND C). AGT acknowledges support from CSIC Research Platform on Quantum Technologies PTI001 and from Spanish project PGC2018-094792-B-100 (MCIU/AEI/FEDER, EU).
\end{acknowledgments}

\appendix

\section{Generation of ancilla-photon superposed states}
\label{apd_sup}
In this appendix, we show that an ancilla-emitter system is further suitable for generating superposed states between the ancilla and the multi-photon MPS [cf.~\cref{mps_sup}]. This is a specific class of the superposed states considered in Ref.~\cite{Miguel-Ramiro2020a}, where many applications of such states are discussed. For example, one can generate a state of the form \cite{Miguel-Ramiro2020a}:
\begin{equation} \label{ghz_cl_sup}
|\psi \rangle  = \frac{1}{{\sqrt 2 }}|0{\rangle _m}|{\rm{GH}}{{\rm{Z}}_{\rm{1}}}\rangle _{\rm ph} + \frac{1}{{\sqrt 2 }}|1{\rangle _m}\left| {{\rm{GH}}{{\rm{Z}}_2}} \right\rangle _{\rm ph},
\end{equation}
where $\left| {{\rm{GH}}{{\rm{Z}}_1}} \right\rangle _{\rm ph} = \frac{1}{{\sqrt 2 }}\left( {\left| {0000} \right\rangle _{\rm ph} - \left| {1111} \right\rangle _{\rm ph}} \right)$ and $\left| {{\rm{GH}}{{\rm{Z}}_2}} \right\rangle  = \frac{1}{{\sqrt 2 }}\left( {\left| {0011} \right\rangle _{\rm ph} + \left| {1100} \right\rangle _{\rm ph}} \right)$. Such a coherent superposition allows one to choose the entanglement structure of the photons by measuring the ancilla in different basis.

To create the states of the form \cref{mps_sup}, let us consider adding an $m$-level ancilla to the MPS source discussed in \cref{mps_prot_gen}. So the system Hilbert space becomes ${{\cal H}_{m}} \otimes {{\cal H}_D} \otimes {{\cal H}_d}$. As pointed out in \cref{mps_prot_gen}, the $m$-level ancilla is not necessarily a physical system. For example, in the system described in the main text (cf.~\cref{sys_mb}), the role of the $m$-level ancilla can be played by a part of the Hilbert space ${{\cal H}_q}$. In this case, one needs $\dim {{\cal H}_q} = 4$ to act as two 2-dimensional ancillas, to create the state of form \cref{ghz_cl_sup}.

Starting from an initial state
\begin{equation} \label{qram_init}
\left| {{\psi _I}} \right\rangle  =  {\sum\limits_{i = 0}^{m - 1} {{\alpha _i}{{\left| i \right\rangle }_{m}}{{\left| {{\varphi _{I,i}}} \right\rangle }_D}} } {\left| 0 \right\rangle _d},
\end{equation}
where ${\left\{ {\left| {{\varphi _{I,i}}} \right\rangle } \right\}_{i = 0,...,D - 1}}$ correspond to the boundary condition of photonic MPS, in the $s$-th round of photon generation we apply the unitary $\hat U\left[ s \right]$ of following form
\begin{equation} \label{}
\hat U\left[ s \right] = \sum\limits_{ijj'kk'} {U_{jk}^{j'k'}} \left[ {i,s} \right]({\left| i \right\rangle _{m}}\langle i|) \otimes ({\left| {j'} \right\rangle _D}\langle j|)({\left| {k'} \right\rangle _d}\langle k|),
\end{equation}
where due to the unitarity of $\hat U\left[ s \right]$,
\begin{equation} \label{unit_cond}
\sum\limits_{p,q} {U_{pq}^{*j'k'}\left[ {i',s} \right]U_{jk}^{pq}\left[ {i,s} \right]}  = {\delta _{ii'}}{\delta _{jj'}}{\delta _{kk'}},\quad \forall s = 1,...,n.
\end{equation}
After the unitary operation, we trigger a photon emission [cf.~\cref{ap_map}]. After applying this sequence for $n$ rounds, the state becomes
\begin{equation} \label{}
\begin{array}{l}
\left| {{\psi _F}} \right\rangle  = {M_P}\hat U\left[ n \right]...{M_P}\hat U\left[ 1 \right]\left| {{\psi _I}} \right\rangle \\
 = \sum\limits_{i = 0}^{m - 1} {{\alpha _i}{{\left| i \right\rangle }_{m}}\left( {U_{\left[ {i,n} \right]}^{{k_n}}...U_{\left[ {i,1} \right]}^{{k_1}}{{\left| {{\varphi _{I,i}}} \right\rangle }_D}} \right){{\left| 0 \right\rangle }_d}} {\left| {{k_n}...{k_1}} \right\rangle _{{\rm{\rm ph}}}},
\end{array}
\end{equation}
where ${\left( {U_{\left[ {i,s} \right]}^{{k_1}}} \right)_{jj'}} \equiv U_{j0}^{j'{k_1}}\left[ {i,s} \right]$. 
In the last step we can disentangle the $D$-level ancilla with the photons and the $m$-level ancilla \cite{schon2005}. For example, we can map the $D$-level ancilla to a state $\left| {{\varphi _F}} \right\rangle_D $ at last. After tracing out the $D$-level ancilla and the $d$-level emitter, we get the desired state [cf.~\cref{mps_sup}] of the $m$-level ancilla and the photons, with
\begin{equation} \label{}
\left| {\psi _{{\rm{MPS}}}^i} \right\rangle _{\rm ph}  = \sum\limits_{{k_n}...{k_1}} {_D\left\langle {{\varphi _F}} \right|U_{\left[ {i,n} \right]}^{{k_n}}...U_{\left[ {i,1} \right]}^{{k_1}}{{\left| {{\varphi _{I,i}}} \right\rangle }_D}{{\left| {{k_n}...{k_1}} \right\rangle }_{{\rm{ph}}}}}.
\end{equation}
Here the unitarity of ${\left\{ {\hat U\left[ s \right]} \right\}_{s = 1,...,n}}$ [cf.~\cref{unit_cond}] lead to the canonical form of ${\left\{ {\left| {\psi _{{\rm{MPS}}}^i} \right\rangle _{\rm ph}} \right\}_{i = 0,...,\left( {m - 1} \right)}}$.

The above process can also be viewed as a quantum random access memory \cite{Giovannetti2008}, where the input quantum address is the initial state \cref{qram_init}, and each (quantum) memory cell is a multi-photon MPS.

%

\section{The quantum gate approach}

\label{circ_unit}

Here, we present the details of the quantum gate approach for generating $D = 2,d = 2$ photonic MPS. As discussed in \cref{stark_circ}, the key step is to implement arbitrary unitaries in ${{\cal H}_{\rm mid}} = {\rm span}( {| 0 \rangle ,| {{1_r}} \rangle ,| {{1_q}} \rangle ,| {{1_r}{1_q}} \rangle } )$. The resonant driving $\mathrm{L_{g}}$ can be decomposed into the $X$ and $Y$ component, with 
\begin{equation} \label{}
	\begin{array}{l}
X\left( \theta  \right) = \exp \left(- {i\left( {{I_q} \otimes {\sigma _x}} \right)\theta /2} \right),\\
Y\left( \theta  \right) = \exp \left(- {i\left( {I_q \otimes {\sigma _y}} \right)\theta /2} \right),
\end{array}
\end{equation}
and we can construct $Z( \theta  ) = X( { - \pi /2} )Y( { - \theta } )X( {\pi /2} )$. On the other hand, the resonant driving $\mathrm{L_{q}}$ can be decomposed into 
\begin{equation} \label{}
\begin{array}{l}
{X_q}\left( \theta  \right) = \exp \left[ { - i\theta \left( {{a_q} \otimes \sigma _r^\dag  + a_q^\dag  \otimes {\sigma _r}} \right)/2} \right],\\
{Y_q}\left( \theta  \right) = \exp \left[ { \theta \left( {a_q^\dag  \otimes {\sigma _r}  - {a_q} \otimes \sigma _r^\dag}  \right)/2} \right],
\end{array}
\end{equation}
which couples the states $\left| {{1_r}} \right\rangle$ and $\left| {{1_q}} \right\rangle$, but also couples $\left| {{1_r}{1_q}} \right\rangle$ to $\left| {{2_q}} \right\rangle$, which is out of ${{\cal H}_{\rm mid}}$. To avoid this leakage, we choose to construct a SWAP gate and the CNOT gate within $\mathcal{H}_{\rm mid}$ without population leakage to state $\left| {{2_q}} \right\rangle $.

Let us apply ${\rm L}_q$ with Rabi frequency ${\Omega _{rq}}$ and a large detuning $\Delta_{\rm ST} \gg {\Omega _{rq}}$. This induces different AC Stark shifts between $\left| {{1_r}} \right\rangle $ and $\left| {{1_q}} \right\rangle $, and between $\left| {{1_r 1_q}} \right\rangle $ and $\left| {{2_q}} \right\rangle $:
 \begin{equation}  
\left| {{1_r}} \right\rangle  \leftrightarrow \left| {{1_q}} \right\rangle :\frac{{\Omega _{rq}^2}}{{2\Delta }},\qquad \left| {{1_r}{1_q}} \right\rangle  \leftrightarrow \left| {{2_q}} \right\rangle :\frac{{\Omega _{rq}^2}}{{\Delta }}.
\end{equation}
By applying the pulse for a time $t = 4{\Delta _{{\text{ST}}}}\theta /\Omega _{{\text{ST}}}^2$, we obtain a phase gate ${S_T}\left( \theta  \right) = {\rm{diag}}\left( {1,{e^{ - i\theta }},{e^{i\theta }},{e^{ - 2i\theta }},{e^{2i\theta }}} \right)$ with the basis $\{ {\left| 0 \right\rangle ,\left| {{1_r}} \right\rangle ,\left| {{1_q}} \right\rangle ,\left| {{1_r}{1_q}} \right\rangle ,\left| {{2_q}} \right\rangle } \}$. Using ${S_T}\left( \theta  \right)$ and resonant laser drivings, the SWAP and CNOT gates are constructed as:
\begin{equation}
\begin{aligned}
{\rm{SWAP}} &= {Y_q}\left( { - \pi /4} \right)Z\left( {\pi /2} \right){S_T}\left( {\pi /2} \right) \hfill \\
&   \cdot {X_q}\left( {\pi /4} \right){S_T}\left( { - \pi /2} \right)Z\left( { - \pi /2} \right), \hfill
\end{aligned} 
\end{equation}
\begin{equation}
{\rm{CNOT = }}X\left( {5\pi /16} \right)Z\left( {\pi /2} \right){S_T}\left( {\pi /2} \right)X\left( {3\pi /16} \right).
\end{equation}
These two gates together with the single-qubit rotations produced by $\mathrm{L}_{\rm g}$ constitute a universal two-qubits gate set in $\mathcal{H}_{\rm mid}$ without leakage error.

We know each unitary in our protocol should act as \cref{u_act}. For the case of generating MPS with $D=2$ and $d=2$, this is equivalent to implementing the isometry ${\hat V_{\left[ i \right]}}:{{\cal H}_{D = 2}} \to {\cal H}_{{\rm{src}}}^{\left( {2,2} \right)}$. One can synthesize a quantum circuit to implement desired isometries, for example using the \textit{UniversalQCompiler} package~\cite{Iten2019}.

As a demonstration, we construct the unitarie that generates the cluster state [\cref{cls_form}]. Using a SVD decomposition approach~\cite{schon2005}, we find two isometries $  {{\hat V}^{\rm cl}_{\left[ {i \ne n} \right]}}$ and $  {{\hat V}^{\rm cl}_{\left[ {n} \right]}}$ needed to generate $n$-photon cluster state
\begin{equation} \label{iso_real}
\begin{gathered}
  {{\hat V}^{\rm cl}_{\left[ {i \ne n} \right]}} = \left( {\begin{array}{*{20}{c}}
  {V_{\left[ i \right]}^0} \\ 
  {V_{\left[ i \right]}^1} 
\end{array}} \right) = \frac{1}{{\sqrt 2 }}\left( {\begin{array}{*{20}{c}}
  1&0 \\ 
  1&0 \\ 
  0&1 \\ 
  0&{ - 1} 
\end{array}} \right),\qquad  \hfill \\
  {{\hat V}^{\rm cl}_{\left[ n \right]}} = \left( {\begin{array}{*{20}{c}}
  {V_{\left[ n \right]}^0} \\ 
  {V_{\left[ n \right]}^1} 
\end{array}} \right) = \left( {\begin{array}{*{20}{c}}
  1&0 \\ 
  0&0 \\ 
  0&1 \\ 
  0&0 
\end{array}} \right). \hfill \\ 
\end{gathered} 
\end{equation}
The corresponding unitaries ${{\hat U}^{\rm cl}_{\left[ {i \ne n} \right]}}$ and ${{\hat U}^{\rm cl}_{\left[ {n} \right]}}$ have the form of \cref{utarg}. The gate sequences for implementing ${{U}_{\left[ i \ne n \right]}}$ and ${{U}_{\left[ n \right]}}$ are shown in Fig.~\ref{cls_circ}, where the gates in the second qubit are implemented with $X,Y,Z$, while the first-qubit rotations are implemented as ${\rm{SWAP}} \to {\textrm{rotations on the second qubit}} \to {\rm{SWAP}}$.

\section{The quantum optimal control (QOC) approach}
\label{goat_imple}
In QOC we aim to find the time-dependence of ${\left\{ {{\Omega _{r\beta} }\left( t \right)} \right\}_{\beta = g,l,q}}$ in $H_{\rm col}(t)$ [\cref{hc_tot}] that yields the desired $U_{\rm targ} \in \mathcal{H}_{\rm col}$ of the form \cref{utarg}, with the embedded isometry ${\hat V_{\rm targ }}$. First, we assume that ${\left\{ {{\Omega _{r\beta} }\left( t \right)} \right\}_{\beta  = g,l,q}}$ are analytical functions parameterized by a set of parameters $\bar \alpha $. The dynamics of our system under the control Hamiltonian $H_{\rm col}(\bar \alpha,t)$ gives rise to the time-ordered $(\mathcal{T}$) evolution operator (with $\hbar  = 1$)
\begin{equation} 
U\left( {\bar \alpha ,T} \right) = {\cal T}\exp \left(- i {\int_0^T {H_{\rm col}(\bar \alpha,t) dt} } \right).
\end{equation}

In numerical calculation, we truncate the dimensions of ${{\cal H}_l}$ and ${{\cal H}_q}$ such that $\dim {U_{\rm targ }} = {N_h}$. The zero matrix $O$ thus is of dimension $\left( {{N_h} - d \cdot D} \right) \times D$. We can denote the unitary generated by our QOC pulse with the same block structure as in \cref{utarg}
\begin{equation}
U\left( {\bar \alpha ,T} \right) = \left( {\begin{array}{*{20}{c}}
{V'\left( {\bar \alpha ,T} \right)}&{{{B'_1}}\left( {\bar \alpha ,T} \right)}\\
{O'\left( {\bar \alpha ,T} \right)}&{{{B'_2}}\left( {\bar \alpha ,T} \right)}
\end{array}} \right).
\end{equation}

We aim to get $V'\left( {\bar \alpha ,T} \right) = {\hat V_{\rm targ}}$ and $O'\left( {\bar \alpha ,T} \right) = O$. Let us define $F_V\left( {\bar \alpha } \right) = {\text{Tr}}\left( {{{\hat V_{\rm targ}}^\dag }V'\left( {\bar \alpha ,T} \right)} \right)$ and $F_O\left( {\bar \alpha } \right) = {\text{Tr}}\left( {O{{\left( {\bar \alpha ,T} \right)}^\dag }O'\left( {\bar \alpha ,T} \right)} \right)$. The deviation of $V'\left( {\bar \alpha ,T} \right)$ from $\hat V_{\rm targ}$ is quantified by $(1 - \left| {{F_V}\left( {\bar \alpha } \right)} \right|/D)$ and the deviation of $O'\left( {\bar \alpha,T } \right)$ from $O$ can be quantified by $\frac{w}{D}\left| {{F_O}\left( {\bar \alpha } \right)} \right|$ with a tunable penalty constant $w$. Thus our optimization task is to minimize the cost function
\begin{equation}
g\left( {\bar \alpha } \right) = 1 - \frac{1}{D}\left| {{F_V}\left( {\bar \alpha } \right)} \right| + \frac{w}{D}\left| {{F_O}\left( {\bar \alpha } \right)} \right|.
\end{equation}

We can evaluate the gradient of $g\left( {\bar \alpha } \right)$ as

\begin{equation} \label{eq:g_grad}
	 \begin{array}{*{20}{l}}
  {{\partial _{\bar \alpha }}g\left( {\bar \alpha } \right) =  - \frac{1}{{D\left| {{F_V}} \right|}}{\text{Re}}\left[ {F_V^* \cdot {\text{Tr}}\left( {\hat V_{{\text{targ}}}^\dag {\partial _{\bar \alpha }}V'} \right)} \right]} \\ 
  { + \frac{w}{{D\left| {{F_O}} \right|}}{\text{Re}}\left[ {F_O^* \cdot {\text{Tr}}\left[ {\left( {{\partial _{\bar \alpha }}{{O'}^\dag }} \right)O' + {{O'}^\dag }{\partial _{\bar \alpha }}O'} \right]} \right],} 
\end{array}
\end{equation}
where ${\partial _{\bar \alpha }}V'\left( {\bar \alpha ,T} \right)$ and ${\partial _{\bar \alpha }}O'\left( {\bar \alpha ,T} \right)$ are the corresponding blocks [cf.~\cref{utarg}] of the matrix ${\partial _{\bar \alpha }}U\left( {\bar \alpha ,T} \right)$. The $U\left( {\bar \alpha ,T} \right)$ and ${\partial _{\bar \alpha }}U\left( {\bar \alpha ,T} \right)$ can be obtained using the GOAT algorithm~\cite{Machnes2018} that numerically integrates the coupled system of the equation of motion
\begin{equation} 
{\partial _t}\left( {\begin{array}{*{20}{c}}
  U \\ 
  {{\partial _{\bar \alpha }}U} 
\end{array}} \right) =  - i\left( {\begin{array}{*{20}{c}}
  H&0 \\ 
  {{\partial _{\bar \alpha }}H}&H 
\end{array}} \right)\left( {\begin{array}{*{20}{c}}
  U \\ 
  {{\partial _{\bar \alpha }}U} 
\end{array}} \right).
\end{equation}
With the knowledge of $g\left( {\bar \alpha } \right)$ and ${\partial _{\bar \alpha }}g\left( {\bar \alpha } \right)$, the control pulse can be optimized by typical gradient descent search methods. In our numerical optimization, we synthesize individual pulses for the real part and the imaginary part for each ${\left\{ {{\Omega _{r\beta} }\left( t \right)} \right\}_{\beta = g,l,q}}$, since this ensures a real value of each pulse amplitude. For each individual component, we choose a Fourier form 
${f}\left( {\bar \alpha ,t} \right) = \sum\nolimits_{j = 1}^{{j_{\max }}} {{A_{j}}\sin \left( {j \cdot {w} \cdot t} \right)}$ as our pulse amplitude, where the parameters are $\bar \alpha  = {\left\{ {{A_{j}},{w}} \right\}_{j = 1,...,{j_{\max }}}}$. We also add two sigmoid functions to bound the amplitude and enforce a smooth start and end of the pulse
\begin{equation}
\begin{gathered}
  {S_1}\left( {f\left( {\bar \alpha ,t} \right)} \right) = b\left( {1 - \frac{2}{{1 + \exp \left( {{g_1} \cdot f\left( {\bar \alpha ,t} \right)/b} \right)}}} \right), \hfill \\
  {S_2}\left( {t,T} \right) = 1 - \frac{2}{{1 + {e^{{g_2}\left( {T - t} \right)}}}}, \hfill \\ 
\end{gathered} 
\end{equation}
where $b,g_1$and $g_2$ are tuning parameters. The overall pulse form for each individual component is $ {S_1}\left( {b,{g_f},f\left( {\bar \alpha ,t} \right)} \right) \cdot {S_2}\left( {t,T} \right)$.

\begin{figure}[!h]
	\centering
\includegraphics[width=0.49\textwidth]{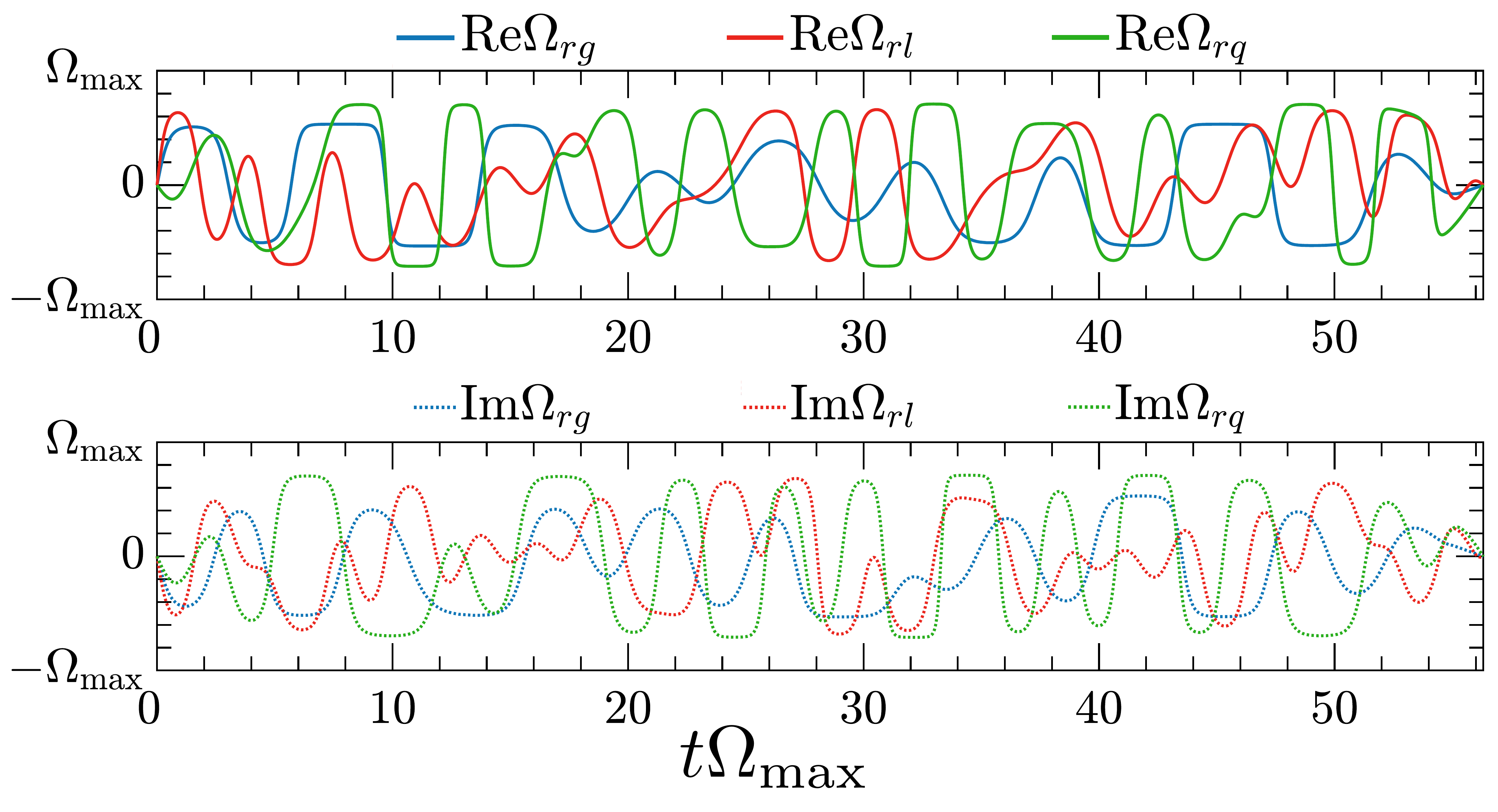}
        \caption{The QOC pulse sequence for implementing ${U_{\left[ {i \ne n} \right]}^{\rm gG}}$ [\cref{u_gghz}] for the generation of the generalized GHZ state of physical dimension $d=3$.}
        \label{gG_pulse}
\end{figure}

As an example, the synthesized pulse for implementing $U_{\left[ {i \ne n} \right]}^{\rm cl}$ (with the corresponding isometry ${\hat V}_{\left[ {i \ne n} \right]}^{\rm cl}$ [\cref{iso_real}]) for the cluster state generation is shown in \cref{cls_circ}, and the pulse for implementing $U_{\left[ {i \ne n} \right]}^{\rm gG}$ for generating the generalized GHZ state \cref{u_gghz} of physical dimension $d=3$ is shown in \cref{gG_pulse}.
We can quantify the quality of the pulses in \cref{cls_circ}(b) and in \cref{gG_pulse} by computing the cost function without the $O$ matrix part as ${g_V}\left( {{U_{{\rm{targ}}}}} \right) = {\left. {g\left( {\bar \alpha } \right)} \right|_{w = 0}}$,
and getting $g_V\left( {U_{\left[ {i \ne n} \right]}^{\rm cl}} \right) = {\rm{1.1}} \times {\rm{1}}{{\rm{0}}^{{\rm{ - 5}}}}$, $g_V\left( {U_{\left[ {i \ne n} \right]}^{\rm gG}} \right) = 2 \times {10^{ - 3}}$, which clearly demonstrate the effectiveness of this method.

\section{The effective modeling of long-time evolution for a Rydberg-blockaded atomic array}
\label{eff_model}
To compute the photonic state fidelity ${{\cal F}_{{\rm{\rm ph}}}} = \left\langle {{\psi _{{\rm{MPS}}}}} \right|{\rho _{{\rm{\rm ph}}}}\left| {{\psi _{{\rm{MPS}}}}} \right\rangle $ efficiently, it will be shown in \cref{mpdo_fid} that we need to numerically obtain the full Liouvillian propagator ${W_{\cal L}}$ [\cref{lv_sol}] for the evolution process, which is a computationally demanding task, as its size grows with the Hilbert space dimension of the atomic array to the power of 4. Thus, directly solving the master equation \cref{ext_me} in the many-body Hilbert space ${\cal H}$ is only possible for array sizes of tens of atoms. To treat the case of larger atom numbers, we need to find an effective master equation that captures the dynamics of the system under the evolution of \cref{ext_me}. For this purpose, we shall first understand the effects of various imperfections discussed in \cref{real_model}.

Recall the basis elements of the collective Hilbert space ${{\cal H}_{\rm col}}$ [cf.~\cref{hcol_basis}]
\begin{equation}
\left| {{s_r}{m_q}{n_l}} \right\rangle  = \frac{1}{{\sqrt {{m_q}!{n_l}!} }}{( {\sigma _r^\dag } )^{{s_R}}}{( {a_q^\dag } )^{{m_q}}}{( {a_l^\dag } )^{{n_l}}}\left| 0 \right\rangle ,
\end{equation}
where $s_r=0,1$ and $m_q,n_l$ are positive integers. First, we include the double Rydberg excitations due to finite van der Waals interaction between Rydberg atoms ${V_{ij}} = {C_6}/r_{ij}^6$. We model its effect with an effectively uniform level shift $U$ for Rydberg doubly excited states, which can be determined from \cref{ushift1}).
Thus, the Rydberg double-excitation can be taken into consideration by adding basis states with $s_r=2$ into ${\mathcal{H}_r}$. Thus now the Rydberg collective excitations form a qutrit with the same properties as a truncated anharmonic oscillator with nonlinearity $U$.

The Rydberg state $\left| {{r}} \right\rangle $ can dephase at a rate ${\Gamma _\phi}$, and can decay to the three hyperfine ground states $\left| g \right\rangle ,\left| q \right\rangle ,\left| l \right\rangle$ with decay rate ${\Gamma _r}$ for each channel. Since the Rydberg decay transition has short wavelength (e.g. $\lambda_{\rm ryd}  = 294$nm in Ref.~\cite{Levine2018}), for a normal $^{87}\mathrm{Rb}$ optical lattice with lattice constant ${d_0} = 532$nm~\cite{Zeiher2015} we have ${d_0}/{\lambda _{\rm ryd}} \approx 1.8$, and therefore the dipole rescattering effect during the Rydberg decay process is very weak.

To get insight for constructing the effective master equation that captures the evolution under \cref{ext_me}, first we analytically solve the decoherence dynamics under \cref{ext_me} with an initial state $\left| {{1_r}} \right\rangle  = a_r^\dag \left| 0 \right\rangle  = \sum\nolimits_{i = 1}^N {{u_i}{e^{i{{\bf{k}}_{rg}} \cdot {{\bf{r}}_i}}}\sigma _{rg}^i} \left| 0 \right\rangle$ by unravelling \cref{ext_me} into a non-Hermitian evolution $S\left( {t,{t_0}} \right)\rho  = {e^{ - i{H_{{\rm{nh}}}}t}} \otimes {e^{i{{\bar H}_{{\rm{nh}}}}t}}\rho $ and the quantum jumps $J$~\cite{Plenio1998} as 
\begin{equation}
\begin{array}{l}
\rho (t) = S\left( {t,{t_0}} \right)\rho \left( {{t_0}} \right)\\
 + \sum\limits_{n = 1}^\infty  {\int_{{t_0}}^t {\rm{d}} } {t_n} \cdots \int_{{t_0}}^{{t_2}} {\rm{d}} {t_1}S\left( {t,{t_n}} \right)J \cdots JS\left( {{t_1},{t_0}} \right)\rho \left( {{t_0}} \right),
\end{array}
\end{equation}
with the non-Hermitian Hamiltonian
\begin{equation} \label{non_hm}
{H_{{\rm{nh}}}} = H'\left( t \right) - \frac{i}{2}\left( {3{\Gamma _r} + {\Gamma _\phi }} \right)\sum\limits_{i = 1}^N {\sigma _{rr}^i} ,
\end{equation}
and the quantum jump given by
\begin{equation}
J\rho  = \sum\limits_{i = 1}^N {\left[ {{\Gamma _r}\sum\limits_{\alpha  = \left\{ {g,q,l} \right\}} {\sigma _{\alpha r}^i \otimes \bar \sigma _{\alpha r}^i}  + {\Gamma _\phi}\sigma _{rr}^i \otimes \bar \sigma _{rr}^i} \right]} \rho,
\end{equation}
where $H'\left( t \right)$ is same as in \cref{ext_me}.
For demonstration, we show the solution under purely spontaneous decay (choosing ${\Gamma _\phi} = 0$) and under purely dephasing (choosing ${\Gamma _{r}} = 0$)
\begin{widetext}
\begin{equation}
\begin{array}{l}
\textrm{purely decay}:\qquad \rho \left( t \right) = \left( {{e^{ - 3{\Gamma _r}t}}a_r^\dag  \otimes \bar a_r^\dag  + \frac{1}{3}\left( {1 - {e^{ - 3{\Gamma _r}t}}} \right)\left( {I + \rho _{q,q}^{\left( 1 \right)} + \rho _{l,l}^{\left( 1 \right)}} \right)} \right)\left| {0 \otimes \bar 0} \right\rangle ,\\
\textrm{purely dephasing}: \qquad \rho \left( t \right) = \left( {{e^{ - {\Gamma _\phi}t}}a_r^\dag  \otimes \bar a_r^\dag  + \left( {1 - {e^{ - {\Gamma _\phi}t}}} \right)\rho _{r,r}^{\left( 1 \right)}} \right)\left| {0 \otimes \bar 0} \right\rangle ,
\end{array}
\end{equation}
\end{widetext}
where we have defined the identity operator $I$ and the density matrix creation operator for singly-excited mixed state $\rho _{\alpha ,\alpha '}^{\left( 1 \right)}$ as
\begin{equation} \label{rhoaa1}
\rho _{\alpha ,\alpha '}^{\left( 1 \right)} = \sum\limits_{i = 1}^N {{{\left| {{u_i}} \right|}^2}\sigma _{\alpha g}^i \otimes \bar \sigma _{\alpha 'g}^i}, \qquad \alpha ,\alpha ' = r,q,l.
\end{equation}
In order to get a simplified description for $\rho _{\alpha ,\alpha '}^{\left( 1 \right)}$, we introduce a set of collective creation operators as 
\begin{equation} \label{mix_ops}
	a_{{M_\alpha }}^\dag  = \sum\limits_{i = 1}^N {{u_i}\sigma _{\alpha g}^i} ,\qquad \alpha  = r,l,q.
\end{equation}
Here ${\{ {a_{{M_\alpha }}^\dag } \}_{\alpha  = l,q}}$ are oscillator creation operators, and $a_{{M_r}}^\dag$ is truncated up to the third Fock state since we include up to double Rydberg excitations. We introduce a projection operator $\mathcal{P}_{\rm diag}$ whose effect is to discard correlation terms with $i \ne j$. Thus we can simplify the description of $\rho _{\alpha ,\alpha '}^{\left( 1 \right)}$ by rewriting it as a projection on the collective excitations as
\begin{equation} \label{proj_rho1}
\begin{array}{l}
\rho _{\alpha ,\alpha '}^{\left( 1 \right)} = {{\cal P}_{{\rm{diag}}}}\left[ {\sum\limits_{i,j = 1}^N {\left( {{u_i}\sigma _{\alpha g}^i} \right) \otimes \left( {{{\bar u}_j}\bar \sigma _{\alpha 'g}^j} \right)} } \right]\\
 = {{\cal P}_{{\rm{diag}}}}\left[ {a_{{M_\alpha }}^\dag  \otimes \bar a_{{M_\alpha }}^\dag } \right].
\end{array}
\end{equation}

Similarly, we can analytically solve the dynamics starting from a doubly excited state $\left| {{2_r}} \right\rangle  = \frac{1}{{\sqrt 2 }}{\left( {a_r^\dag } \right)^2}\left| 0 \right\rangle$. The solution for the purely decay case (${\Gamma _\phi} = 0$) is $\rho \left( t \right) = {\rho ^{\left( 0 \right)}}\left( t \right) + {\rho ^{\left( 1 \right)}}\left( t \right) + {\rho ^{\left( 2 \right)}}\left( t \right)$, with 
\begin{widetext}
\begin{equation} \label{ex2_dm}
	\begin{array}{l}
{\rho ^{\left( 0 \right)}}\left( t \right) = {e^{ - 6{\Gamma _r}t}} \frac{1}{{\sqrt 2 }}{\left( {a_r^\dag } \right)^2} \otimes {\left( {\bar a_r^\dag } \right)^2}\left| {0 \otimes \bar 0} \right\rangle, \\
{\rho ^{\left( 1 \right)}}\left( t \right) = \frac{2}{3}\left( {{e^{ - 3{\Gamma _r}t}} - {e^{ - 6{\Gamma _r}t}}} \right)\left( {a_r^\dag  \otimes \bar a_r^\dag } \right) \cdot \left[ {I + \rho _{q,q}^{\left( 1 \right)} + \rho _{l,l}^{\left( 1 \right)}} \right]\left| {0 \otimes \bar 0} \right\rangle, \\
{\rho ^{\left( 2 \right)}}\left( t \right) = \frac{1}{9}{\left( {1 - {e^{ - 3{\Gamma _r}t}}} \right)^2}\left( {{\rho ^{\left( 0 \right)}} + 2\rho _{q,q}^{\left( 1 \right)} + 2\rho _{l,l}^{\left( 1 \right)} + 2\rho _{ql,ql}^{\left( 2 \right)} + \rho _{qq,qq}^{\left( 2 \right)} + \rho _{ll,ll}^{\left( 2 \right)}} \right)\left| {0 \otimes \bar 0} \right\rangle,
\end{array}
\end{equation}
\end{widetext}
where we defined a general notation 
\begin{equation} \label{rmix_comp}
\rho _{\left\{ \alpha  \right\},\left\{ {\alpha '} \right\}}^{\left( {{n_e}} \right)} = \sum\limits_{{i_1} \ne ... \ne {i_{{n_e}}} = 1}^N {\prod\limits_{k = 1}^{{n_e}} {{{\left| {{u_{{i_k}}}} \right|}^2}\sigma _{{\alpha _k}g}^{{i_k}} \otimes \bar \sigma _{\alpha'_k g}^{{i_k}}} } .
\end{equation}
Here $\left\{ \alpha  \right\} \equiv {\alpha _1}...{\alpha _{{n_e}}},\left\{ {\alpha }' \right\} \equiv \alpha_1 ' ...\alpha _{{n_e}}'$ and ${\alpha _i},\alpha _i' = r,l,q$. Similar to \cref{proj_rho1}, \cref{rmix_comp} can be rewritten using the $\mathcal{P}_{\rm diag}$ as
\begin{equation} \label{rho_cp}
\rho _{\left\{ \alpha  \right\},\left\{ {\alpha '} \right\}}^{\left( {{n_e}} \right)} = {{\cal P}_{{\rm{diag}}}}\left[ {\prod\limits_{k = 1}^{{n_e}} {a_{{M_{{\alpha _k}}}}^\dag  \otimes \bar a_{{M_{\alpha'_k}}}^\dag } } \right].
\end{equation}
Here the effect of $\mathcal{P}_{\rm diag}$ is to discard correlation terms where $\left\{ {{i_1},...,{i_{n_e}}} \right\} \ne \left\{ {{j_1},...,{j_{n_e}}} \right\}$. 

We can also solve the pure-dephasing (${\Gamma _r} = 0$) dynamics starting from $\left| {{2_r}} \right\rangle $ as
\begin{equation} \label{dep_dyn}
\begin{aligned}
  \rho \left( t \right) &= \left[{e^{ - 2{\Gamma _\phi }t}}\frac{1}{{\sqrt 2 }}{\left( {a_r^\dag } \right)^2} \otimes {\left( {\bar a_r^\dag } \right)^2} \hfill \right. \\
&   + 2t{e^{ - 2{\Gamma _\phi }t}}\left( {a_r^\dag  \otimes \bar a_r^\dag } \right) \cdot \rho _{r,r}^{\left( 1 \right)} \hfill \\
  &\left. + \left( {1 - {e^{ - 2{\Gamma _\phi }t}} - 2t{e^{ - 2{\Gamma _\phi }t}}} \right)\rho _{rr,rr}^{\left( 2 \right)}\right] \left| {0 \otimes \bar 0} \right\rangle . \hfill \\ 
\end{aligned} 
\end{equation}
Thus we see that during our sequential MPS generation protocol, the dephasing of $\left| r \right\rangle $ and the decay from $\left| r \right\rangle $ to $\left| l \right\rangle $ and $\left| q \right\rangle $ produce mixed excitations, and these mixed states with $n_e$-excitations are represented by 
\begin{equation} 
{{\rho _{\rm mix}} = {\cal N}_{\left\{ \alpha  \right\},\left\{ {\alpha '} \right\}}\rho _{\left\{ \alpha  \right\},\left\{ {\alpha '} \right\}}^{\left( {{n_e}} \right)}\left| {0 \otimes \bar 0} \right\rangle },	
\end{equation}
where ${\cal N}_{\left\{ \alpha  \right\},\left\{ {\alpha '} \right\}}$ is a normalization factor.

Eq.~(\ref{rho_cp}) allows us to greatly simplify the description of $\rho_{\rm mix}$ since now it is characterized by its number of collective excitations. To carefully understand the evolution of ${\rho _{\rm mix}}$ under the master equation \cref{ext_me}, we study its non-Hermitian evolution and the quantum jump one by one. Defining the non-Hermitian evolution operator as 
\begin{equation}
{U_{\rm nh}}\left( t \right) = {\cal T}\exp \left( { - i\int_0^t {{H_{\rm nh}}\left( \tau  \right)d\tau } } \right),
\end{equation}
the non-Hermitian evolution of $\rho _{\rm mix}$ is 
\begin{equation} \label{Unh_evo}
\begin{array}{*{20}{l}}
  {{U_{\rm nh}}\left( t \right) \otimes {{\bar U}_{nh}}\left( t \right){\rho _{{\text{mix}}}} = {U_{\rm nh}}\left( t \right) \otimes {{\bar U}_{nh}}\left( t \right)} \\ \times {\cal N}_{\left\{ \alpha  \right\},\left\{ {\alpha '} \right\}}{  {{\cal P}_{{\text{diag}}}}\left[ {\prod\limits_{k = 1}^{{n_e}} {a_{{M_{{\alpha _k}}}}^\dag  \otimes \bar a_{{M_{{\alpha' _{k}}}}}^\dag } } \right]\left| {0 \otimes \bar 0} \right\rangle .} 
\end{array}
\end{equation}
The effect of ${U_{\rm nh}}\left( t \right)$ can be understood on the single-operator level,
\begin{equation}
\begin{array}{*{20}{l}}
{{U_{{\rm{nh}}}}\left( t \right)\sigma _{{\alpha _1}g}^\dag ...\sigma _{{\alpha _n}g}^\dag \left| 0 \right\rangle  = \sigma _{{\alpha _1}g}^\dag \left( t \right)...\sigma _{{\alpha _n}g}^\dag \left( t \right){U_{{\rm{nh}}}}\left( t \right)\left| 0 \right\rangle ,}\\
{{\text{with}}\quad \sigma _{\alpha g}^\dag \left( t \right) = {U_{{\rm{nh}}}}\left( t \right)\sigma _{\alpha g}^\dag U_{\rm nh}^{ - 1}\left( t \right).}
\end{array}
\end{equation}
When $H_{\rm nh}(t)$ does not include the driving $\mathrm{L_{g}}$, we always have ${U_{\rm nh}}\left( t \right)\left| 0 \right\rangle  = \left| 0 \right\rangle $. Also notice that $H_{\rm nh}(t)$ is a collection of single-particle Hamiltonians ${H_{\rm nh}}(t) = \sum\nolimits_{i = 1}^N {{H_{i}}}(t) $, such that $\sigma _{\alpha'g}^\dag$ with $\alpha' \ne \alpha $ does not get mixed into $\sigma _{\alpha g} ^\dag \left( t \right)$. Thus the order of $\mathcal{P}_{\rm diag}$ and $U_{\rm nh}\left( t \right)$ in \cref{Unh_evo} can be exchanged as
\begin{equation} \label{coh_evo}
\begin{array}{*{20}{l}}
{{U_{\rm nh}}\left( t \right) \otimes {{\bar U}_{nh}}\left( t \right){{\cal P}_{{\rm{diag}}}}\left[ {\prod\limits_{k = 1}^{{n_e}} {a_{{M_{{\alpha _k}}}}^\dag  \otimes \bar a_{{M_{{\alpha' _{k}}}}}^\dag } } \right]\left| {0 \otimes \bar 0} \right\rangle }\\
{ = {{\cal P}_{{\rm{diag}}}}\left[ {{U_{\rm nh}}\left( t \right) \otimes {{\bar U}_{nh}}\left( t \right)\prod\limits_{k = 1}^{{n_e}} {a_{{M_{{\alpha _k}}}}^\dag  \otimes \bar a_{{M_{{\alpha' _{k}}}}}^\dag } } \right]\left| {0 \otimes \bar 0} \right\rangle .}
\end{array}
\end{equation}

When $\mathrm{L}_g$ is turned on, the non-Hermitian dynamics becomes more involved. We approach it from a different perspective, by looking at the purity $P_r$ of ${\rho _{\rm mix}}$. One gets
\begin{equation} 
P_r=\mathrm{Tr}\left[ {{\rho^2 _{\rm mix}}} \right] = {\cal N}_{\left\{ \alpha  \right\},\left\{ {\alpha '} \right\}}\sum\limits_{{i_1} \ne ... \ne {i_{n_e}}} {{{\left| {{u_{{i_1}}}} \right|}^4}...{{\left| {{u_{{i_{{n_e}}}}}} \right|}^4}}.
\end{equation}
In our scheme, we consider $\mathrm{L}_g$ to be a laser beam that shines on the whole array. Thus its beam profile $u_i$ in general scales as $u_i\sim O\left( {1/\sqrt N } \right)$, and ${\cal N}_{\left\{ \alpha  \right\},\left\{ {\alpha '} \right\}}\sim O\left( {1/{n_e}!} \right)$. Thus ${{P_r}}\sim O\left(1/ {{n_e}!{N^{{n_e}}}} \right) \to 0$ when $N \gg 1$. On the other hand, we know that unitary evolution does not change $P_r$. Thus, starting from $\rho_{\rm mix}$, when the system is only driven by $\mathrm{L}_g$, for an array with $N \gg 1$ the purity is always $P_r\approx 0$. The consequence is that Rydberg mixed excitation(s) almost do not evolve back to $\left| g \right\rangle $ under $\mathrm{L}_g$. Thus, in our effective model we make the approximation that ${\rho _{\rm mix}}$ does not change under $\mathrm{L}_g$ driving.

Similarly to the non-Hermitian evolution case, the action of quantum jump $J$ on ${\rho _{\rm mix}}$ can also be simplified by changing the order of $J$ and the projection $\mathcal{P}_{\rm diag}$, since $J$ is also a collection of single atom jumps
\begin{equation}\label{jump_mix}
J{\rho _{{\rm{mix}}}} = {\cal N}_{\left\{ \alpha  \right\},\left\{ {\alpha '} \right\}}{{\cal P}_{{\rm{diag}}}}\left[ {J\prod\limits_{k = 1}^{{n_e}} {a_{{M_{{\alpha _k}}}}^\dag  \otimes \bar a_{{M_{\alpha' {_k}}}}^\dag } } \right]\left| {0 \otimes \bar 0} \right\rangle .
\end{equation}

Thus, the evolution of ${\rho _{\rm mix}}$ can be obtained by a projection $\mathcal{P}_{\rm diag}$ on the Lindblad dynamics of collective excitations $\{ {a_{M_\alpha}^\dag } \}$[cf.~\cref{mix_ops}], as shown in \cref{coh_evo} and \cref{jump_mix}. The Fock space of these collective excitations expands a Hilbert space ${{\cal H}_{\rm mix}}$. 

\begin{figure}[h!]
	\centering
	\includegraphics[width=0.49\textwidth]{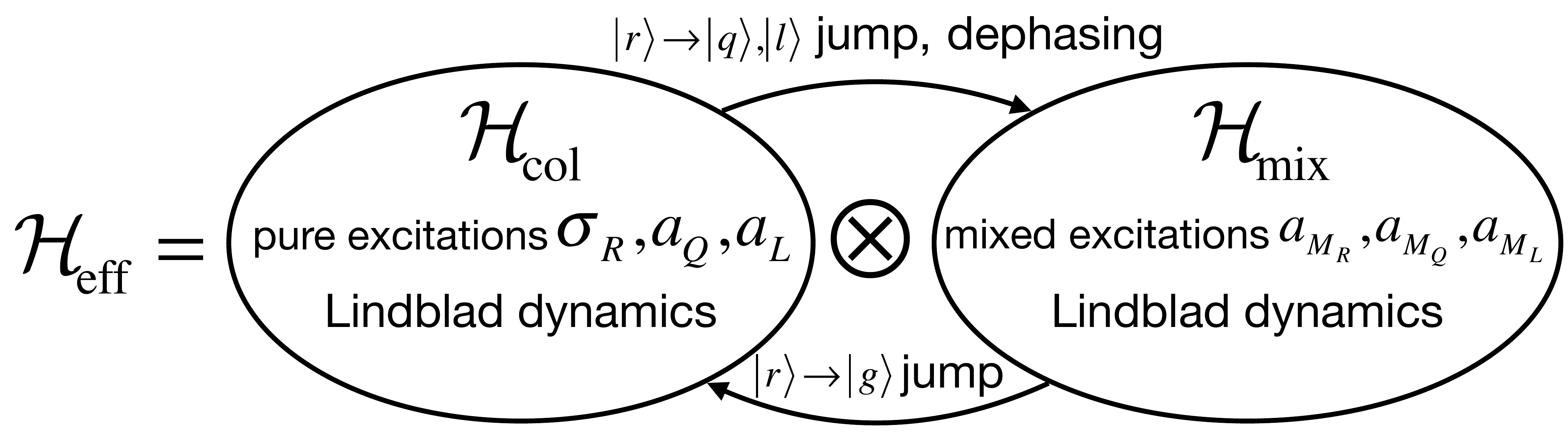}
        \caption{Illustration of the coupling structure for $\mathcal{H}_{\rm eff}$.}
        \label{heff}
\end{figure}

Together with the Lindblad evolution of collective excitations in $\mathcal{H}_{\rm col}$, the Hilbert space of our whole effective model is ${{\cal H}_{\rm eff}} = {{\cal P}_{\rm Ryd}}\left[ {{\cal H}_{\rm col}} \otimes {{\cal H}_{\rm mix}}\right]$. Here, ${{\cal P}_{\rm Ryd}}$ projects out the states with more than two Rydberg excitations. The unitary evolution preserves the population within ${{\cal H}_{\rm col}}$ and ${{\cal H}_{\rm mix}}$, while the Rydberg decoherence processes couple these two spaces. This structure is illustrated in \cref{heff}.

The system density matrix ${\rho _{\rm eff}}(t)\in {\mathcal{H}_{{\text{\rm eff}}}}$ evolves under the following Lindblad master equation
\begin{equation} \label{lindblad}
\begin{gathered}
  {{\dot \rho }_{\rm eff}}\left( t \right) =  - i\left[ {H_{\rm eff}\left( t \right),{\rho _{\rm eff}}\left( t \right)} \right] \hfill \\
   + \sum\limits_n {\frac{{1}}{2}\left[ {2{C_n}{\rho _{\rm eff}}\left( t \right)C_n^\dag  - {\rho _{\rm eff}}\left( t \right)C_n^\dag {C_n} - C_n^\dag {C_n}{\rho _{\rm eff}}\left( t \right)} \right]} , \hfill \\ 
\end{gathered} 
\end{equation}
with Hamiltonian
\begin{equation} \label{heff_tot}
{H_{{\rm{eff}}}}\left( t \right) = {{\cal P}_{{\rm{Ryd}}}}\left[ {H_0^{{\rm{eff}}} + H_{rg}^{{\rm{eff}}}\left( t \right) + H_{rq}^{{\rm{eff}}}\left( t \right) + H_{rl}^{{\rm{eff}}}\left( t \right)} \right].
\end{equation}
Defining ${\omega _{{M_\alpha }}} = {\omega _\alpha }$ with $\alpha  \in \left\{ {r,q,l} \right\}$, we have
\begin{equation}
{H_0^{{\rm{eff}}}} = \sum\limits_{\alpha  = r,l,q,{M_r},{M_l},{M_q}} {{\omega _\alpha }} a_\alpha ^\dag {a_\alpha } + \frac{U}{2}{\hat n_r} \cdot \left( {{{\hat n}_r} - 1} \right),
\end{equation}
with ${\hat n_r} = a_r^\dag {a_r} + a_{{M_r}}^\dag {a_{{M_r}}}$, and
\begin{equation}
    {H^{{\rm{eff}}}_{rg}}\left( t \right) = \frac{{{\Omega _{rg}}\left( t \right)}}{2}\left( {{e^{ - i{\omega _{rg}}t}}a _r^\dag  + h(c).} \right),
\end{equation}
\begin{equation}
\begin{aligned}
{H^{{\rm{eff}}}_{r\alpha }}\left( t \right) &= \frac{{{\Omega _{r\alpha }}\left( t \right)}}{2}\left[{e^{ - i{\omega _{r\alpha }}t}}( {a_r} \otimes a_\alpha ^\dag \right. \\
&\left.+ {a_{{M_r}}} \otimes a_{{M_\alpha }}^\dag  ) + {\rm H.c.}\right], \quad \alpha  = l,q.
\end{aligned}
\end{equation}

Following the result of the analytical solution of \cref{ex2_dm}, we can write down the jump operators in \cref{lindblad} for the Rydberg decay process as
\begin{equation} 
\begin{array}{*{20}{l}}
{{C_{R \to G}} = \sqrt {{\Gamma _r}} {a_r},\qquad {C_{{M_r} \to G}} = \sqrt {{\Gamma _r}} {a_{{M_r}}},}\\
{C_{R \to {M_\alpha }}} = \sqrt {{\Gamma _r}} {a_r}\sigma _{{M_\alpha }}^\dag ,\qquad {C_{{M_r} \to {M_\alpha }}} = \sqrt {{\Gamma _r}} {a_{{M_r}}}\sigma _{{M_\alpha }}^\dag ,\\
\mathrm{with} \quad \alpha  = l,q,
\end{array}
\end{equation}
where the $\sigma _{{M_\alpha }}^\dag $ for $\alpha  = r,l,q$ have same structure as oscillator creation operators, but without the $\sqrt{n}$ dependence. The jump operators for the Rydberg dephasing process are
\begin{equation} 
    {C_{R\to{M_r}}} = \sqrt {{\Gamma _\phi}} {a _r}\sigma _{{M_r}}^\dag ,\qquad{C_{{M_r}\to{M_r}}} = \sqrt {{\Gamma _\phi}} \sigma _{\phi},
\end{equation}
where ${\sigma _{\phi}} = {\rm diag}\left( {0,1,\sqrt 2 } \right)$ to capture the evolution in \cref{dep_dyn}.

\subsection*{Benchmark with exact simulation}

\begin{figure}[h!]
    \centering
    \includegraphics[width=0.49\textwidth]{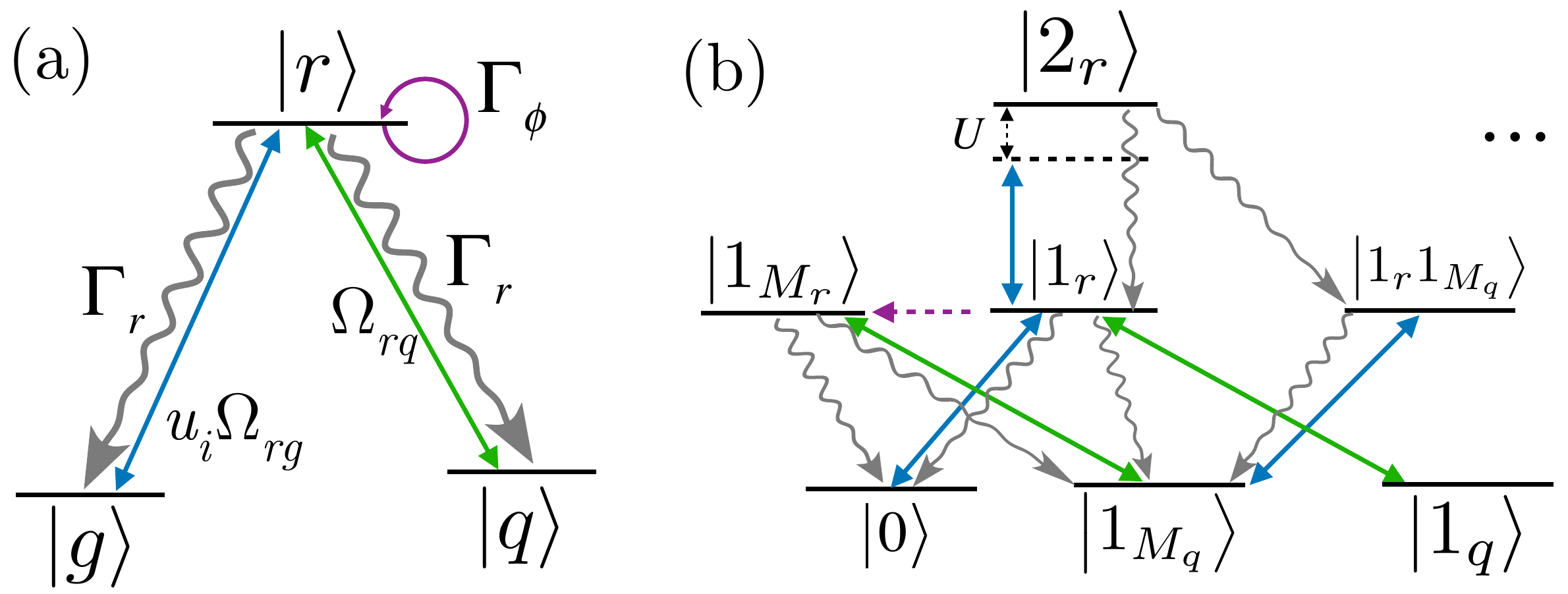}
        \caption{The simplified atom level structure for the benchmark. (a) The single-atom level scheme. (b) A part of the level scheme for the collective effective model. The basis is defined in \cref{error_hilbert_basis}. The two-sided arrows correspond to the laser couplings in (a) with the same coloring. The purple arrow corresponds to the dephasing channel.}
        \label{raman_level}
\end{figure}

We provide a benchmark of our effective modeling by comparing its prediction with that obtained by numerically solving the master equation \cref{ext_me}. In order to numerically reach larger system size for solving \cref{ext_me}, we consider a simplified atomic-level structure that only involves states $\left| g \right\rangle ,\left| q \right\rangle ,\left| r \right\rangle $ as shown in \cref{raman_level}(a), and include all errors discussed in this section. The benchmark on this simplified level structure is enough to demonstrate the validity of our effective modeling since the state $\left| q \right\rangle $ and $\left| l \right\rangle $ experience the same errors.

We test our effective model on a cyclic Raman transition process between the state $\left| 0 \right\rangle $ and $\left| {{1_q}} \right\rangle $, which the evolution starts from state $\left| 0 \right\rangle $ and going through following process 
\begin{equation} \label{rabi_cyc}
\left| 0 \right\rangle  \to \left| {{1_r}} \right\rangle  \to \left| {{1_q}} \right\rangle  \to \left| {{1_r}} \right\rangle  \to \left| 0 \right\rangle \to...    
\end{equation}
back and forth. Such a population transfer is very common in our protocol. 
We count each transfer process between $\left| 0 \right\rangle $ and $\left| {{1_q}} \right\rangle $ as one time of Raman transfer. The errors appear when the Rydberg state $\left| r \right\rangle $ gets populated during each Raman transfer, thus this cyclic process allows us to see how the errors accumulate in the long-time evolution.

Specifically, for this cyclic Raman transfer process we alternatively apply $\pi $ pulses with resonant laser coupling ${{\rm{L}}_{\rm{g}}}$ and ${{\rm{L}}_{\rm{q}}}$ of constant Rabi frequencies ${\Omega _{rg}}$ and ${\Omega _{rq}}$, with coupling time ${t_{rg}} = \pi /{\Omega _{rg}}$ and ${t_{rq}} = \pi /{\Omega _{rq}}$.
For simplicity, here we consider the beam profile of ${{\rm{L}}_{\rm{g}}}$ as a plane wave, such that ${u_i} = 1/\sqrt N $ in \cref{exci_op_g}. In the numerical calculation, we truncate the many-body Hilbert space ${{\cal H}}$ up to two excitations, which is enough to simulate the case of the photonic MPS generation with $D=2$ and $d=2$. In this case, we have 
\begin{equation}
\begin{aligned}
{\cal H} = {\rm{span}}( {\left| g \right\rangle ^{ \otimes N}} ,{{\left\{ {\left| {{q_i}} \right\rangle ,\left| {{r_i}} \right\rangle } \right\}}_{i = 1\sim N}},\\{{\left\{ {\left| {{r_i}{q_j}} \right\rangle ,\left| {{q_i}{r_j}} \right\rangle ,\left| {{r_i}{r_j}} \right\rangle ,\left| {{q_i}{q_j}} \right\rangle } \right\}}_{i < j = 1\sim N}} ),
\end{aligned}
\end{equation}
with $\dim {{\cal H}} = 2{N^2} + 1$. 

The basis of ${{\cal H}_{\rm eff}}$ for the simplified level structure before applying the projection ${{\cal P}_{{\rm{diag}}}}$ is:
\begin{equation} \label{error_hilbert_basis}
\begin{array}{l}
\left| {{s_r}{m_q},{{\tilde s}_{{M_r}}}{{\tilde m}_{{M_q}}}} \right\rangle  = \\
\sqrt {\dfrac{{{{\cal N}_{\left\{ \alpha  \right\},\left\{ \alpha  \right\}}}}}{{{s_r}!{m_q}!}}} {\left( {a_r^\dag } \right)^{{s_r}}}{\left( {a_q^\dag } \right)^{{m_q}}}{\left( {a_{{M_r}}^\dag } \right)^{{{\tilde s}_{{M_r}}}}}{\left( {a_{{M_q}}^\dag } \right)^{{{\tilde m}_{{M_q}}}}}\left| 0 \right\rangle .
\end{array}
\end{equation}

\begin{figure*}
    \centering
    \includegraphics[width=\textwidth]{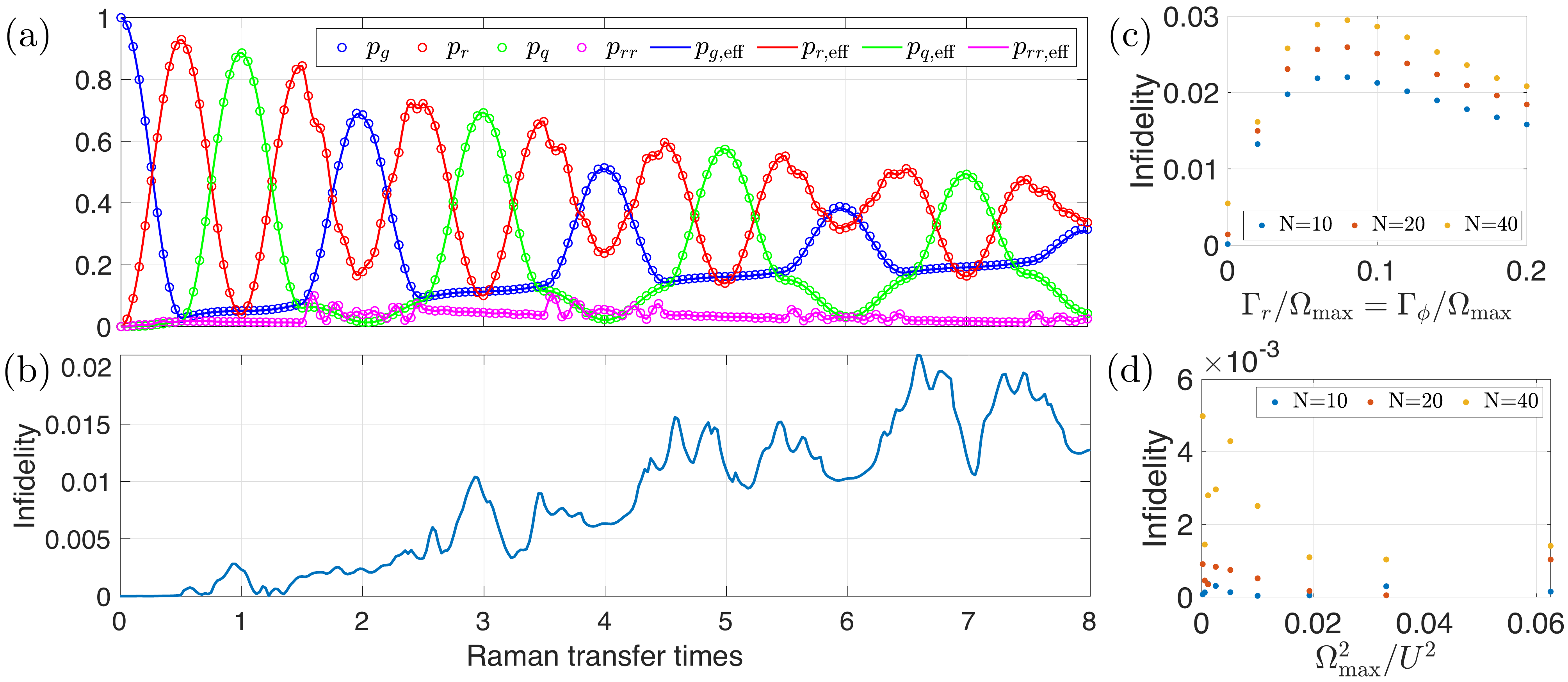}
        \caption{Benchmark of the effective model with a cyclic Raman transfer process. (a) The time evolution of the ratios of total population with no excitation ($p_g$), with one excitation on $\left| q \right\rangle $ ($p_q$), with one Rydberg excitation ($p_r$) and with two Rydberg excitations ($p_{rr}$) over 8 Raman transfers. The rounded points are exact solution, and the lines correspond to the results from the effective model simulation. (b)  The evolution of the infidelity for the same process as in (a). (c) The infidelity after 10 Raman transfers under various imperfections rates. (d) The infidelity after 10 Raman transfers for various Rydberg nonlinear shifts $U$.}
        \label{eff_fid}
\end{figure*}
In \cref{raman_level}(b), we pictorially demonstrate the coupling between a part of the basis in ${\cal H}_{\rm eff}$ during this cyclic Raman transfer process.

For demonstration, in \cref{eff_fid}(a) we compare the system dynamics obtained by solving the effective master equation [\cref{lindblad}] with the exact solution obtained by solving the exact master equation [\cref{ext_me}] for the following choice of parameters:
\begin{equation} 
\begin{array}{l}
N = 20,\qquad {\Gamma _r}/{\Omega _{\max }} = 0.016,\qquad \\
{\Gamma _\phi }/{\Omega _{\max }} = 0.016,\qquad U/{\Omega _{\max }} = 5.
\end{array}
\end{equation}
We see a good agreement between the exact soluation and prediction of the effective model. To quantify the validity of the effective model, we compare these two dynamics on the full density matrix level. The exact master equation simulation provides us with the time-dependent density matrix ${\rho }\left( t \right) \in {\mathcal{H}}$, while the effective model provides another time-dependent density matrix ${\rho _{\rm eff}}\left( t \right) \in {\mathcal{H}_{{\text{eff}}}}$. To directly compare ${\rho _{\rm eff}}\left( t \right)$ and ${\rho}\left( t \right)$, we reconstruct a density matrix in ${{\cal H}}$ from ${\rho _{\rm eff}}\left( t \right)$ by mapping the corresponding basis from ${{\cal{H}}_{\rm eff}}$ to ${{\cal H}}$. For example
\begin{equation} 
\begin{array}{l}
 \left| {{1_r}} \right\rangle \left\langle {{1_r}} \right| \to \sum\limits_{i,j = 1}^N {{u_i}u_j^*{e^{i{{\bf{k}}_{rg}} \cdot \left( {{{\bf{r}}_i} - {{\bf{r}}_j}} \right)}}} \left| {{r_i}} \right\rangle \left\langle {{r_j}} \right|, \\
\mathrm{and}\quad \left| {{1_{{M_r}}}} \right\rangle \left\langle {{1_{{M_r}}}} \right| \to \sum\limits_{i,j = 1}^N {{{\left| {{u_i}} \right|}^2}} \left| {{r_i}} \right\rangle \left\langle {{r_i}} \right|.
\end{array}
\end{equation}
From this mapping we obtain a density matrix $\rho _{{\rm{eff}}}^{{\rm{mb}}}(t) \in {\cal H}$ predicted by our effective model. Then we can compute the infidelity $1 - {F_{{\rm{eff}}}}\left( t \right)$ between $\rho _{{\rm{eff}}}^{{\rm{mb}}}(t) \in {\cal H}$ and ${\rho}(t)$, where
\begin{equation} 
{F_{{\rm{eff}}}}\left( t \right) = {\left( {{\rm{Tr}}\sqrt {\sqrt {\rho \left( t \right)} \rho _{{\rm{eff}}}^{{\rm{mb}}}\left( t \right)\sqrt {\rho \left( t \right)} } } \right)^2}.
\end{equation}
The evolution of the infidelity for the dynamics in \cref{eff_fid}(a) is shown in \cref{eff_fid}(b). During the entire evolution that contains 8 Raman transfers, the infidelity stays very low. We further obtain the infidelity after 10 Raman transfers for various parameters, which is shown in \cref{eff_fid}(c-d). In all the regimes that our protocol runs (see the parameter range in \cref{fid_scale}), we see good agreement between the exact simulation and the prediction from our effective model. Thus this effective modeling reliably captures the dynamics of the atomic array over a long time.
\section{Multi-photon emission and the construction of the process map $W_P$ for photon emission process}
\label{multi_HP_calc}
With a finite retrieval efficiency $p_{\rm em}$, the photon emission process of $\left| {{1_l}} \right\rangle $ is
\begin{equation} 
\left| {{1_l}} \right\rangle  \xrightarrow{{{M_P}}} \left| 0 \right\rangle \left( {\sqrt {{p_{{\rm{em}}}}} {{\left| 1 \right\rangle }_{{\rm{\rm ph}}}}\left| 0 \right\rangle _{\epsilon _{{\text{\rm ph}}}} + \sqrt {1 - {p_{{\rm{em}}}}} {{\left| 0 \right\rangle }_{{\rm{\rm ph}}}}\left| 1 \right\rangle _{\epsilon _{{\text{\rm ph}}}}} \right),
\end{equation}
Where ${\epsilon _{{\text{\rm ph}}}}$ is a photonic environmental mode used to capture the erroneous jump. Under the Holstein-Primakoff approximation [\cref{ops_approxi}], the multiphoton emission property for a state $\left| {{m_q}{n_l}} \right\rangle $ can be obtained directly from the single-photon emission property~\cite{Porras2008}, which gives
\begin{equation} 
\begin{gathered}
  \left| {{m_q}{n_l}} \right\rangle \xrightarrow{{{M_P}}} \hfill \\
  \sum\limits_{j = 0}^{{n_l}} {\sqrt {{\binom{n_l}{j}}p_{{\text{em}}}^j{{\left( {1 - {p_{{\text{em}}}}} \right)}^{{n_l} - j}}} } \left| {{m_q}} \right\rangle {\left| j \right\rangle _{{\text{\rm ph}}}}\left| {{n_l} - j} \right\rangle _{\epsilon _{{\text{\rm ph}}}}. \hfill \\ 
\end{gathered} 
\end{equation}
\subsection*{Error due to the Holstein-Primakoff approximation}
In real experiments there will be an additional photon retrieval error due to the deviation of the collective excited states in many-body Hilbert space $\cal H$ from that obtained under the Holstein-Primakoff approximation [\cref{hcol_basis}]. For simplicity, let us consider the collective excitation profile being close to a plane wave as ${u_i} \approx 1/\sqrt N$. In this case the exact form of ${\left| {{m_q}{n_l}} \right\rangle _{\rm ext}}$ in $\cal H$ and ${\left| {{m_q}{n_l}} \right\rangle}$ are
\begin{equation} \label{multi_state}
\begin{gathered}
\begin{aligned}
  {\left| {{m_q}{n_l}} \right\rangle _{{\text{ext}}}} = &\sqrt {\frac{{{N^{{m_q} + {n_l}}}\left( {N - {m_q} - {n_l}} \right)!}}{{{m_q}!{n_l}!N!}}}  \hfill \\
  & \cdot {\left( {S_{q,{{\mathbf{k}}_{rg}} - {{\mathbf{k}}_{rq}}}^\dag } \right)^{{m_q}}}{\left( {S_{l,{{\mathbf{k}}_{rg}} - {{\mathbf{k}}_{rl}}}^\dag } \right)^{{n_l}}}\left| 0 \right\rangle , 
\end{aligned} \hfill \\
  \left| {{m_q}{n_l}} \right\rangle  = \frac{1}{{\sqrt {{m_q}!{n_l}!} }}{\left( {a_q^\dag } \right)^{{m_q}}}{\left( {a_l^\dag } \right)^{{n_l}}}\left| 0 \right\rangle.  \hfill \\ 
\end{gathered} 
\end{equation}
Approximately expanding the operators $S$ and $a$ in \cref{multi_state} both using spin operators [cf.~\cref{exci_op_g}], we can compute the overlap of these two states in the regime of ${m_q} + {n_l} \ll N$ as
\begin{equation} 
\left\langle {{m_q}{n_l}} \right.{\left| {{m_q}{n_l}} \right\rangle _{{\text{ext}}}} \approx \sqrt {1 - \frac{{\left( {{m_q} + {n_l}} \right)\left( {{m_q} + {n_l} - 1} \right)}}{{2N}}} .
\end{equation}
To estimate the effect that this difference has on photon retrieval, we decompose $\left| {{m_q}{n_l}} \right\rangle $ as 
\begin{equation} 
\begin{aligned}
  {\left| {{m_q}{n_l}} \right\rangle _{{\text{ext}}}} &\approx \sqrt {1 - \frac{{\left( {{m_q} + {n_l}} \right)\left( {{m_q} + {n_l} - 1} \right)}}{{2N}}} \left| {{m_q}{n_l}} \right\rangle  \hfill \\
&   + \sqrt {\frac{{\left( {{m_q} + {n_l}} \right)\left( {{m_q} + {n_l} - 1} \right)}}{{2N}}} {\left| \phi  \right\rangle _{{\text{err}}}}, \hfill \\ 
\end{aligned} 
\end{equation}
where ${\left| \phi  \right\rangle _{\rm err}}$ is a state that is orthogonal to ${\left| {{m_q}{n_l}} \right\rangle}$. We overestimate the photon retrieval error by assuming that ${\left| \phi  \right\rangle _{\rm err}}$ does not generate a directional photon that can be retrieved. This lead to an state-dependent retrieval efficiency
\begin{equation} \label{}
{p_{{\rm{em}}}}\left( {\left| {{m_q}{n_l}} \right\rangle_{\rm ext} } \right) = \left( {1 - \frac{{\left( {{m_q} + {n_l}} \right)\left( {{m_q} + {n_l} - 1} \right)}}{{2N}}} \right)p_{{\rm{em}}}^{{n_l}}.
\end{equation}
For simplicity, we represent this state dependence by a renormalized
single photon emission rate $p'_{\rm em}$ for ${\left| {{m_q}{n_l}} \right\rangle }_{\rm ext}$ that scales as
\begin{equation} 
{p'_{{\rm{em}}}}\left( {{{\left| {{m_q}{n_l}} \right\rangle }_{{\rm{ext}}}}} \right) = {\left( {1 - \frac{{\left( {{m_q} + {n_l}} \right)\left( {{m_q} + {n_l} - 1} \right)}}{{2N}}} \right)^{1/{n_l}}}{p_{{\rm{em}}}}.
\end{equation}
We further overestimate the multi-excitation retrieval error by determining ${p'_{{\rm{em}}}}$ from largest excitation number, which relates to the bond dimension $D$ and physical dimension $d$ of the photonic state that we want to generate as follows:
\begin{equation} 
\max {m_q} = D - 1, \quad \max {n_l} = d - 1.
\end{equation}
Thus we get an lower bound of the retrieval efficiency [cf.~\cref{reno_pem}]
\begin{equation} 
{p'_{{\rm{em}}}} \ge {\left( {1 - \frac{{\left( {D + d - 2} \right)\left( {D + d - 3} \right)}}{2N}} \right)^{1/\left( {d - 1} \right)}}{p_{{\rm{em}}}}.
\end{equation}
For example in the cluster state generation, the state with maximal possible excitation number is $\left| {{1_q}{1_l}} \right\rangle$, with $D = 2$ and $d = 2$. Thus for the cluster state generation, we get a reduced retrieval efficiency  ${p'_{{\rm{em}}}} \ge \left( {1 - \frac{1}{N}} \right){p_{{\rm{em}}}}$.
\subsection*{The construction of $W_P$}
Now we can construct the the process map ${W_P}:{\mathcal{H}_{{\text{eff}}}} \to {\mathcal{H}_{{\text{eff}}}} \otimes {\mathcal{H}_{{\text{\rm ph}}}}$ for the photon emission process. In the effective modeling [\cref{eff_model}], the general basis of $\mathcal H _{\rm eff}$ is
\begin{widetext}
\begin{equation}
\left| {{s_r}{m_q}{n_l},{{\tilde s}_{{M_r}}}{{\tilde m}_{{M_q}}}{{\tilde n}_{{M_l}}}} \right\rangle  = \sqrt {\frac{{{{\cal N}_{\left\{ \alpha  \right\},\left\{ \alpha  \right\}}}}}{{{s_r}!{m_q}!{n_l}!}}} {\left( {a_r^\dag } \right)^{{s_r}}}{\left( {a_q^\dag } \right)^{{m_q}}}{\left( {a_l^\dag } \right)^{{n_l}}}{\left( {a_{{M_r}}^\dag } \right)^{{{\tilde s}_{{M_r}}}}}{\left( {a_{{M_q}}^\dag } \right)^{{{\tilde m}_{{M_q}}}}}{\left( {a_{{M_l}}^\dag } \right)^{{{\tilde n}_{{M_l}}}}}\left| 0 \right\rangle,
\end{equation}
with ${{s_r} + {{\tilde s}_{{M_r}}} \leqslant 2}$ since we include at most two Rydberg excitations. By introducing three additional environmental modes ${\epsilon _r},{\epsilon _{{M_r}}},{\epsilon _{{M_l}}}$, we can denote the emission process ${M_P}:{\mathcal{H}_{{\text{eff}}}} \to {\mathcal{H}_{{\text{eff}}}} \otimes {\mathcal{H}_{{\text{\rm ph}}}}$ as
\begin{equation} \label{mp_emit}
{M_P}:\left| {{s_r}{m_q}{n_l},{{\tilde s}_{{M_r}}}{{\tilde m}_{{M_q}}}{{\tilde n}_{{M_l}}}} \right\rangle  \to \sum\limits_{j = 0}^{{n_l}} {\sqrt {{{{\binom{n_l}{j}}\left( {{{p}'_{{\text{em}}}}} \right)^j}}{{\left( {1 - {{p}'_{{\text{em}}}}} \right)}^{{n_l} - j}}} } \left| {{m_q},{{\tilde m}_{{M_q}}}} \right\rangle {\left| j \right\rangle _{{\text{\rm ph}}}}\left| {{n_l} - j} \right\rangle _{\epsilon _{{\text{\rm ph}}}}{\left| {{s_r}} \right\rangle _{{\epsilon _r}}}{\left| {{{\tilde s}_{{M_r}}}} \right\rangle _{{\epsilon _{{M_r}}}}}{\left| {{{\tilde n}_{{M_l}}}} \right\rangle _{{\epsilon _{{M_l}}}}}.
\end{equation}
\end{widetext}
Here we assumed that all excitations on $\left| l \right\rangle $ and $\left| r \right\rangle $ completely decays away after the photon emission process. From \cref{mp_emit} we can construct $W_P$ as 
\begin{equation} \label{wp_map}
{W_P} = {\rm Tr}{_{{\epsilon _{{\text{\rm ph}}}},{\epsilon _r},{\epsilon _{{M_r}}},{\epsilon _{{M_l}}}}}\left[ {{M_P} \otimes {{\bar M}_P}} \right].	
\end{equation}

\section{Computing photonic state fidelity using the matrix product density operator (MPDO) approach}
\label{mpdo_fid}
In this section we provide the formalism to compute the fidelity ${{\cal F}_{\rm ph}} = {\left\langle {{\psi _{\rm MPS}}} \right|{\rho _{\rm ph}}\left| {{\psi _{\rm MPS}}} \right\rangle } $ for generating $\left| {{\psi _{\rm MPS}}} \right\rangle $ [\cref{mps}] with Rydberg-blockaded atomic array [cf. \cref{seq_gen}], where ${\rho _{\rm ph}}$ is the eventually created photonic density matrix in our protocol including all errors. Let us consider the effective modeling of the atomic array [cf. \cref{eff_model}] with the density matrix ${\rho _{{\text{eff}}}}(t) \in {\mathcal{H}_{{\text{eff}}}}$. The ${\rho _{\rm ph}}$ is determined by the dynamics of ${\rho _{\rm eff}}(t)$, which follows \cref{lindblad}. Denote ${N_{{\text{eff}}}} \equiv \dim {\mathcal{H}_{{\text{eff}}}}$, \cref{lindblad} can be written as a vectorized form
\begin{equation}\label{rho_v}
\frac{{d\vec \rho_{\rm eff} \left( t \right)}}{{dt}} = \mathcal{L}\left( t \right)\vec \rho_{\rm eff} \left( t \right), \quad{\vec \rho _{\rm eff}}(t) = \sum\limits_{a,b = 0}^{{N_{\rm eff}-1}} {{\rho_{ab}}}(t) \left| {a \otimes \bar b} \right\rangle.
\end{equation}
Here $\mathcal{L}\left( t \right)$ is the Liouville operator, and ${\bar b}$ represent the complex conjugate of $b$. The solution of \cref{rho_v} is 
\begin{equation} \label{lv_sol}
{\vec \rho _{\rm eff}}\left( T \right) = \mathcal{T}\left\{ {{e^{\int_0^T {\mathcal{L}\left( t \right)dt} }}} \right\}{\vec \rho _{\rm eff}}\left( 0 \right) = {W_{\cal L}}{\vec \rho _{\rm eff}}\left( 0 \right),
\end{equation}
where the Liouvillian propagator ${W_{\cal L}}$ is of dimension $\dim \left({{W_{\cal L}}} \right) = N_{\rm eff}^4$.

The process map  ${W_P}$ [\cref{wp_map}] for the photon emission process can be written as
\begin{equation} 
{W_P} = \sum\limits_{i,j = 0}^{d_{\rm max} - 1} {\sum\limits_{a,b,c,s = 0}^{{N_{\rm eff}} - 1} {W_{P,abcs}^{ij}} } \left| {c,\bar s,i,\bar j} \right\rangle \left| {a,\bar b} \right\rangle ,
\end{equation}
which maps the system density matrix with vectorized basis $| {a \otimes \bar b} \rangle $ to an source--photon joint density matrix with vectorized basis $| {c,\bar s,i,\bar j} \rangle  = | {c \otimes \bar s} \rangle  \otimes {| {i \otimes \bar j} \rangle _{{\text{\rm ph}}}}$. Here $d_{\rm max}-1>d-1$ denote the maximal photon number that can be generated in the emission process, although the $d'-1>d-1$ levels are not used in the computation of ${\cal F}_{\rm ph}$.

 Thus in total, each round of photon generation consisting of an evolution under the master equation [\cref{lindblad}] followed by the photon emission $W_P$, results in a map
\begin{equation}
\begin{aligned}
  {{\vec \rho }_{\left[ k \right]}} &= \sum\limits_{{i_k},{j_k} = 0}^{d_{\rm max} - 1} {N_{\left[ k \right]}^{{i_k},{j_k}}{{\vec \rho }_{\left[ {k - 1} \right]}}} ,  \hfill \\ N_{\left[ k \right]}^{{i_k},{j_k}} &= W_P^{{i_k}{j_k}}{W_{{{\cal L}_{\left[ k \right]}}}}. \hfill \\ 
\end{aligned} 
\end{equation}
\newline
Starting from an initial state $\left| {{\varphi _I}} \right\rangle  \in {\mathcal{H}_{{\text{eff}}}}$ and successively operate the protocol for $n$ rounds, we arrive at a source--photon density matrix
\begin{equation}
{\vec \rho _{\left[ n \right]}} = \sum\limits_{\left\{ {{i_k},{j_k}} \right\} = 0}^{d_{\rm max}-1} {N_{\left[ n \right]}^{{i_n},{j_n}}...N_{\left[ 1 \right]}^{{i_1},{j_1}}\left| {{\varphi _I},{i_n}...{i_1}} \right\rangle \left| {{\varphi _I},{j_n}...{j_1}} \right\rangle }.
\end{equation}
The $n$-photon density matrix ${\rho _{\rm ph}}$ is obtained by tracing out the source
\begin{equation}
{\vec \rho _\mathrm{\rm ph}} = \sum\limits_{\left\{ {{i_k},{j_k}} \right\} = 0}^{d_{\rm max}-1} {{\text{Tr}}\left[ {N_{\left[ n \right]}^{{i_n},{j_n}}...N_{\left[ 1 \right]}^{{i_1},{j_1}}\tilde B} \right]\left| {{i_n},...,{i_1}} \right\rangle _{\rm ph} \left| {{j_n},...,{j_1}} \right\rangle _{\rm ph}}  ,
\end{equation}
with $\tilde B = \sum\nolimits_{a  = 0}^{N_{\rm eff}-1} {\left| {{\varphi _I}} \right\rangle \langle a | \otimes \left| {{\varphi _I}} \right\rangle \langle a |} $. Defining $B = \left| {{\varphi _I}} \right\rangle \left\langle {{\varphi _F}} \right|$, the fidelity ${{\cal F}_{\rm ph}}$ can then be efficiently evaluated as
\begin{equation} \label{fid_exp}
\begin{aligned}
{{\cal F}_{{\rm{\rm ph}}}} &= \sum\limits_{\left\{ {{i_k},{j_k}} \right\} = 0}^{d - 1} {{\rm{Tr}}\left[ {N_{\left[ n \right]}^{{i_n},{j_n}}...N_{\left[ 1 \right]}^{{i_1},{j_1}}\tilde B} \right]} \\
& \times {\rm{Tr}}\left[ {\left( {V_{\left[ n \right]}^{{j_n}} \otimes \bar V_{\left[ n \right]}^{{i_n}}} \right)...\left( {V_{\left[ 1 \right]}^{{j_1}} \otimes \bar V_{\left[ 1 \right]}^{{i_1}}} \right)\left( {B \otimes B} \right)} \right].
\end{aligned}
\end{equation}

\section{photon retrieval process}

\label{apd_retri}
In this section, we present the details for calculating the photon retrieval efficiency ${p_{\rm em}}$. Our calculations in this section follow Ref.~\cite{Manzoni2018} closely.

As explained in \cref{pho_ret_main}, we consider the photon emission from the singly-excited initial state [\cref{em_init}], and the decay dynamics can be described by an ansatz $\left| {\psi _{\rm em}\left( t \right)} \right\rangle  = \left( {{l_j}\left( t \right)\sigma _{lg }^j + {e_j}\left( t \right)\sigma _{eg}^j} \right)\left| 0 \right\rangle $ which evolves under a non-Hermitian Hamiltonian ${H_{\rm em}}(t) = {H_{el}}(t) + {H_{{\rm{DDI}}}}$ [\cref{nonh_ham}].  In our calculation we set the polarizations of all atoms along $x$ direction to maximize the photon emission speed along $z$ direction, that ${ {\bf{d}}_j} = { {\bf{x}}} \forall j$ [cf.~\cref{retri_results}(a)]. We get the equation of motion as
\begin{equation} \label{atom_evo}
\begin{array}{l}
{{\dot e}_j} = {\rm{i}}\Delta {e_j} - {\rm{i}}\dfrac{{{\Omega _{{el}}}(t)}}{2}{l_j} + {\rm{i}}{\Gamma _{\rm em}}\sum\limits_l {{M_{jl}}} {e_l},\\
{{\dot l}_j} =  - {\rm{i}}\dfrac{{{\Omega _{{el}}}(t)}}{2}{e_j},
\end{array}	
\end{equation}
where ${\Gamma _{{\rm{em}}}} = {\mu _0}\omega _{eg}^3d_{eg}^2/3\pi \hbar c$ is the spontaneous emission rate from $\left| e \right\rangle $ to $\left| g \right\rangle $ and ${M_{jl}} = 3\pi k_0^{ - 1} {\bf{d}}_j^* \cdot {{\bf{G}}_0}\left( {{{\bf{r}}_j},{{\bf{r}}_l},{\omega _{eg}}} \right) \cdot {{\bf{d}}_l}$. \cref{atom_evo} can be further simplified by noticing that ${M_{jl}}$ is a symmetric complex matrix, thus can be transpose-diagonalized~\cite{Asenjo-Garcia2017}, and we will obtain a set of eigenvalues $\left\{ {\lambda _\xi } \right\}$ and corresponding eigenmodes $\left\{ {{{\mathbf{v}}_\xi }} \right\}$ that satisfies ${\bf{v}}_\xi ^T \cdot {{\bf{v}}_{{\xi ^\prime }}} = {\delta _{\xi {\xi ^\prime }}}$ and $\sum\nolimits_\xi  {{{\bf{v}}_\xi }}  \otimes {\bf{v}}_\xi ^T = {\bf{I}}$.

By solving atomic state evolution [\cref{atom_evo}] and relating it to the emitted photonic field via an input-output relation~\cite{Dung2002,Caneva2015,Manzoni2018}, one can compute the photon retrieval efficiency ${p_{\rm em}}$ as the probability to go to a defined detection mode ${{\bf E}_{\det }}\left( r \right)$. One can define a detection mode operator~\cite{Manzoni2018}
\begin{equation} 
	{\hat E_{{\rm{det}}}} = {\rm{i}}{d_{eg}}\sqrt {\frac{{{k_0}}}{{2 {\epsilon _0}{F_{{\rm{det}}}}}}} \sum\limits_j {{\bf E}_{{\rm{det}}}^*} \left( {{{\bf{r}}_j}} \right) \cdot {{\bf d}_j}\sigma _{ge}^j,
\end{equation}
where the factor ${F_{{\rm{det }}}} = \int_{z = {\rm{ const }}} {{{\rm{d}}^2}} {\bf{r}}{\bf{E}}_{{\rm{det}}}^*({\bf{r}}) \cdot {{\bf{E}}_{{\rm{det}}}}({\bf{r}})$ normalizes ${\hat E_{{\rm{det}}}}$ such that $\left\langle {\hat E_{{\rm{det}}}^\dag {{\hat E}_{{\rm{det}}}}} \right\rangle $ represents the photon number per unit time emitted into the detection mode. The retrieval efficiency ${p_{\rm em}}$ can be represented as
\begin{equation} 
{p_{\rm em}} = \int_0^\infty  {\rm{d}} t\left\langle {\hat E_{{\rm{det}}}^\dag (t){{\hat E}_{{\rm{det}}}}(t)} \right\rangle .
\end{equation}
By solving the dynamics of $\left| {\psi _{\rm em}\left( t \right)} \right\rangle$, one can show~\cite{Manzoni2018}
\begin{equation} \label{pem_form}
{p_{{\rm{em}}}} = \dfrac{{{S_{{\lambda _{eg}}}}}}{{4{F_{{\rm{det}}}}}}\sum\limits_{j,l}^N {{u_j}{K_{jl}}u_l^*},
\end{equation}
where ${S_{{\lambda _{eg}}}} = (3/2\pi )\lambda _{eg}^2$ is the resonant atomic optical cross-section. The matrix $K$ has a form  
\begin{equation} \label{}
{K_{jl}} = {\rm{i}}\sum\limits_{\xi ,{\xi ^\prime } = 1}^N {{v_{\xi ,j}}v_{{\xi ^\prime },l}^*\frac{{E_\xi ^*{E_{{\xi ^\prime }}}}}{{{\lambda _\xi } - \lambda _{{\xi ^\prime }}^*}}} ,
\end{equation}
where ${E_\xi } = \sum\nolimits_m {{v_{\xi ,j}}} {E_j}$ and ${E_j} = {{\bf{E}}_{{\rm{det}}}}\left( {{{\bf{r}}_j}} \right) \cdot {\bf{d}}_j^*$.

As introduced in the main text, we consider two retrieval schemes: a uni-directional retrieval scheme where the photon goes to a single direction, and a two-directional retrieval scheme where the photon goes to two opposite directions (shown in \cref{retri_results}). For the uni-directional retrieval scheme, we set the detection mode as a single vector Gaussian mode~\cite{Chen2002a}
\begin{equation} 
\begin{gathered}
  E_{{\text{det}}}^x({\mathbf{r}}) = {E_0}\int_0^1 {\text{d}} bb{{\text{e}}^{ - {b^2}k_0^2w_0^2/4}}{{\text{e}}^{{\text{i}}{k_0}z\sqrt {1 - {b^2}} }}{J_0}\left( {b{k_0}\rho } \right), \hfill \\
E_{{\rm{det}}}^z({\bf{r}}) =  - {\rm{i}}{E_0}\frac{x}{\rho }\int_0^1 {\rm{d}} b\frac{{{b^2}{J_1}\left( {b{k_0}\rho } \right)}}{{\sqrt {1 - {b^2}} }} \cdot {{\rm{e}}^{ - {b^2}k_0^2w_0^2/4 + {\rm{i}}{k_0}z\sqrt {1 - {b^2}} }}, \hfill \\ 
\end{gathered} 
\end{equation}
where ${J_0}$ and ${J_1}$ are the Bessel functions. For the two-directional retrieval scheme, we choose a symmetric superposition of two vector Gaussian modes from two opposite directions as our detection mode. 

The smallest photon retrieval error $\varepsilon _{\rm em}^{\rm opt}$ corresponds to the largest eigenvalue of the matrix $K$ in \cref{pem_form}, with the corresponding eigenvector as its excitation profile, denoted as the optimal excitation profile. We also study the case of initial atomic excitation profile as a Gaussian mode ${l_j}\left( 0 \right) \propto E_{\det }^x\left( {{{\bf{r}}_j}} \right)$, which in principle can be easily created with the laser $\rm {L_g}$ and $\rm {L_l}$, where $\rm {L_g}$ has the Gaussian profile and $\rm {L_l}$ has plane-wave profile. The retrieval error in this case is denoted as $\varepsilon _{\rm em}^{\rm Gauss}$, as shown in \cref{pho_ret_main}.

\subsection*{Effect of array defects and atomic thermal motion}
\label{ret_imperf}

\begin{figure}[h!]

	\centering
	\includegraphics[width=0.48\textwidth]{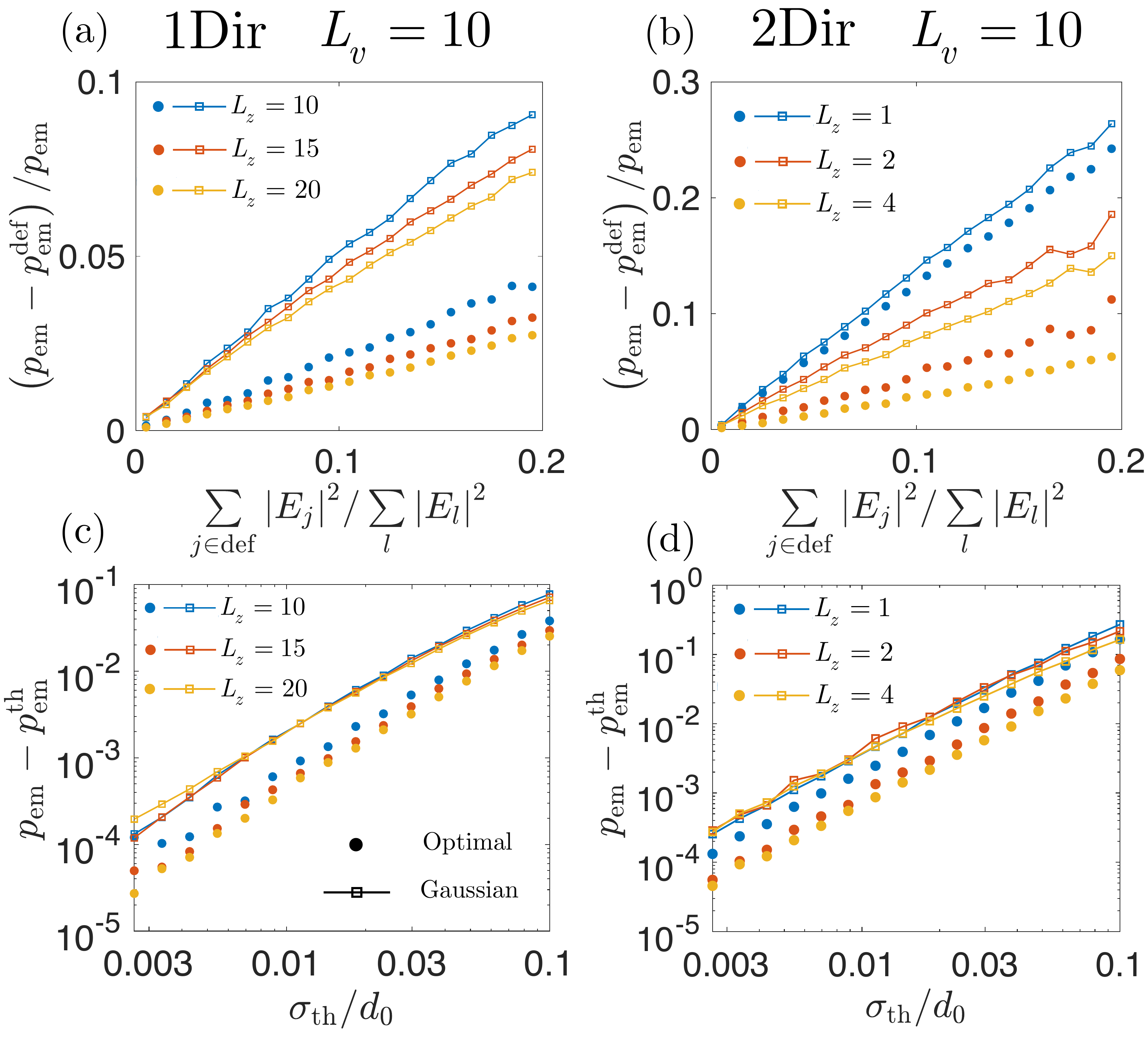}
        \caption{The effect of atomic defects and thermal motion on the photon retrieval efficiency $p_{\rm em}$. (a) The relative difference between the retrieval efficiency for perfect array $p_{\rm em}$ and efficiency $p^{\rm def}_{\rm em}$ of array with defects on random positions, as a function of $\sum\nolimits_{j \in \rm def} {{{\left| {{E_j}} \right|}^2}/\sum\nolimits_l {{{\left| {{E_l}} \right|}^2}} }$ for uni-directional retrieval scheme. Here each point is obtained by averaging evenly separated intervals of in $x$ axis. (b) The same as (a), but for two-directional retrieval scheme. (c) and (d) The difference between $p_{\rm em}$ and the retrieval efficiency $p^{\rm th}_{\rm em}$ with atomic positional disorder as a function of standard deviation $\sigma_{\rm th}$ of the positional disorder. Each point is averaged over 50 realizations of atomic positions with the same ${\sigma _{\rm th}}/d_0$.}
        \label{disorder_results}
\end{figure}

To model the effect of atomic defects, we generate random atomic configurations where a fraction (up to 20\%) of positions in the array are unoccupied. For each fraction, we generate 100 realizations of the configuration. As shown in Ref.~\cite{Manzoni2018}, one expects that the relative decrease of the retrieval efficiency should be proportional to the ratio of the detection mode hitting the empty sites. The results for the uni-directional (two-directional) retrieval is shown in \cref{disorder_results}(a) (\cref{disorder_results}(b)), where we separate a series of the intervals of $\sum\nolimits_{j \in \rm def} {{{\left| {{E_j}} \right|}^2}/\sum\nolimits_l {{{\left| {{E_l}} \right|}^2}} }$, and average the photon retrieval errors within each interval. From these data we obtain the scaling of the modification of $p_{\rm em}$ due to atomic defects as in \cref{p_def}. we also see that $\alpha_{\rm def}$ decreases with the increase of array size, showing that an array with a larger size is more robust against this error. These results are consistent with Ref.~\cite{Manzoni2018}.

To model the effect of atomic thermal motion, we add random spatial disorder $\delta {\mathbf{r}_j} = \left( {{\delta _{x,j}},{\delta _{y ,j}},{\delta _{z,j}}} \right)$ where each component is randomly distributed with a standard deviation $\sigma_{\rm th}$. The error of the effect of photon retrieval efficiency are shown in \cref{disorder_results}(c) and \cref{disorder_results}(d), with each point obtained by averaging over 50 realizations of atomic positions with the same ${\sigma _{\rm th}}/d_0$. In those figures, we see that the effect of this random disorder is proportional of ${\sigma_{\rm th} ^2}/{d_0^2}$, which is the same as predicted in Refs.~\cite{Manzoni2018, Shahmoon2017}.

\section{Scaling of the achievable photon number with the bond and physical dimension of MPS}
\label{dim_scale}

As pointed out in the introduction, MPS with moderate high bond and physical dimensions already find many applications, and can well capture the ground states of 1D local gapped Hamiltonians \cite{Huang2015a,Dalzell2019,Huang2019a,Schuch2017}. Even for one-dimensional spin models at a critical point, a moderately high $D$ is enough to capture the ground state when this chain only have moderate system size, as the main deviation to the thermodynamic limit result comes from the finite size effect, instead of a finite-entanglement effect provided by the limited $D$ \cite{pirvu2012matrix}. In this section, we provide a simple estimation of the scaling of the maximum achievable photon number as a function of the bond dimension and the physical dimension of MPS. 

The photon number $N_{\rm ph}$ is determined by the error rate per photon as \cref{err_opt_mid}. To produce an MPS with bond dimension $D$ and physical dimension $d$, we need to implement a unitary on the Hilbert space of dimension $2 \cdot \dim {\cal H}_{src}^{\left( {D,d} \right)} = 2Dd$. There is numerical evidence suggesting that \cite{Lee2018b} the time cost ${T_{D,d}}$ of implementing a general unitary in such a Hilbert space using the quantum optimal control approach scales as ${T_{D,d}} \sim {\left( {Dd} \right)^2}$. Thus, it takes more time to implement unitaries when we want to produce photonic MPS with higher bond and physical dimensions, which lead to stronger decoherence as the coefficients ${\beta _U},{\beta _r },{\beta _\phi}$ in \cref{err_opt_mid} are proportional to ${T_{D,d}}$. Also the retrieval efficiency is bounded by \cref{reno_pem}. In the regime of $D + d \ll N, {\beta _0} \approx 0$ and ${p_{{\rm{em}}}} \approx 1$, the worst case estimation yields the following scaling of error
\begin{equation} \label{scale_err_D}
\begin{aligned}
\xi _{{\rm{opt}}}^{D,d} & \approx {T_{D,d}}{\left( {\dfrac{{27{\beta _U}}}{{4f_{{L_v},{L_z}}^2}}} \right)^{1/3}}{\left( {\dfrac{{\Gamma d_0^6}}{{\left| {{C_6}} \right|}}} \right)^{2/3}}\\
& - {\beta _{{\rm{em}}}}\dfrac{{(D + d - 2)(D + d - 3)}}{{2\left( {d - 1} \right)N}}.
\end{aligned}
\end{equation}
Thus the dominant part lead to a qualitative scaling of ${N_{{\rm{\rm ph}}}}\sim{D^{ - 2}}{d^{ - 2}}$.

Note that there are ways to reduce the error. As discussed in \cref{tot_pho_opt}, one can use higher Rydberg levels to improve the non-linearity and add a cavity to reduce the photon retrieval error.


\bibliography{./library.bib}
\end{document}